%% file: webfp-v2.tex
\documentclass[twocolumn]{svjour3}
\usepackage[]{microtype}
\usepackage{numberedblock}
\usepackage{url}
\usepackage{amsfonts,amsmath,amssymb}

\usepackage{subfig}
\usepackage{multirow}
\usepackage{multicol}
\usepackage{floatrow}
\usepackage{paralist}
\usepackage{graphicx}
\usepackage{float}
\usepackage{csvsimple}
\usepackage{rotating}
\usepackage{pdflscape}
\usepackage{booktabs}
\usepackage{cite}

\usepackage{algorithm}
\usepackage{algorithmic}
\floatname{algorithm}{Framework}

\newcommand{\ra}[1]{\renewcommand{\arraystretch}{#1}}

\graphicspath{{./figures/}}

\newcommand{\ie}{{i.e.}}
\newcommand{\eg}{{e.g.}}
\newcommand{\etal}{et al.}

\newcommand{\mahdiC}[1]{{\color{red}Comment by Mahdi:} {\color{blue}#1}}
\newcommand{\ahC}[1]{{\color{red}Comment by Amirhossein:} {\color{blue}#1}}

\newcommand{\amiraliC}[1]{{\color{red}Comment by Amirali:} {\color{blue}#1}}

\newcommand{\drMJC}[1]{{\color{red}Comment by Dr:} {\color{blue}#1}}

\journalname{ArXiv}

\usepackage{subfig}
\usepackage{multirow}
\usepackage{floatrow}

\usepackage{tikz}
\usepackage{csvsimple,booktabs,siunitx}
\usepackage[numbers]{natbib}

\usepackage[inline]{enumitem}

\newcolumntype{L}[1]{>{\raggedright\arraybackslash}p{#1}}
\newcolumntype{C}[1]{>{\centering\arraybackslash}p{#1}}
\newcolumntype{R}[1]{>{\raggedleft\arraybackslash}p{#1}}

\begin{document}
\title{Machine Learning Interpretability Meets TLS Fingerprinting}

\author{Mahdi Jafari Siavoshani, Amirhossein Khajehpour$^*$, Amirmohammad Ziaei$^*$, Amirali Gatmiri, Ali Taheri\\
	Information, Network, and Learning Lab (INL)\\
	Computer Science and Engineering Department\\
	Sharif University of Technology, Tehran, Iran
\institute{
 Mahdi~Jafari~Siavoshani \at Sharif University of Technology, Tehran, Iran \\ \email{mjafari@sharif.edu} \and
 Amirhossein~Khajehpour \at Sharif University of Technology, Tehran, Iran \\ \email{khajepour.amirhossein@gmail.com} \and
 Amirmohammad~Ziaei \at Sharif University of Technology, Tehran, Iran \\ \email{amirzia76@gmail.com} \and
 Amirali~Gatmiri \at Sharif University of Technology, Tehran, Iran \\ \email{gatmiri@ce.sharif.edu} \and
 Ali~Taheri \at Sharif University of Technology, Tehran, Iran \\ \email{ali.taheri3745@gmail.com} \and
 $^*$These authors have equal contributions to this work.
}
}

%




\date{Received: date / Accepted: date}
\authorrunning{Jafari Siavoshani \etal}
\maketitle

\begin{abstract}
Protecting users' privacy over the Internet is of great importance; however, it becomes harder and harder to maintain due to the increasing complexity of network protocols and components. Therefore, investigating and understanding how data is leaked from the information transmission platforms and protocols can lead us to a more secure environment. 
	
In this paper, we propose a framework to systematically find the most vulnerable information fields in a network protocol. To this end, focusing on the transport layer security (TLS) protocol, we perform different machine-learning-based fingerprinting attacks on the collected data from more than 70 domains (websites) to understand how and where this information leakage occurs in the TLS protocol. Then, by employing the interpretation techniques developed in the machine learning community and applying our framework, we find the most vulnerable information fields in the TLS protocol. Our findings demonstrate that the TLS handshake (which is mainly unencrypted), the TLS record length appearing in the TLS application data header, and the initialization vector (IV) field are among the most critical leaker parts in this protocol, respectively.

\end{abstract}

	\section{Introduction} \label{sec:Introduction}

\input{01-introduction}
	
	\section{Related Works}\label{sec:relatedWorks}
	\input{02-related-works}

	\section{Backgrounds on Machine Learning Methods}\label{sec:MLMethods}
	\input{03-machine-learning-methods}\label{sec:Background}

	\section{Dataset}\label{sec:dataset}
	\input{04-dataset}

	\section{Methodology}\label{sec:Methodology}
	\input{05-methodology}

	\section{Experimental Results}\label{sec:results}
	\input{06-experiment}

	
	\section{Conclusion and Future Works}\label{sec:conc}
	\input{08-conclusion}

	\section*{Acknowledgment}
	\input{09-acknowledge}

	\bibliographystyle{spbasic}	
	\bibliography{webfp-ref}

	
	\appendix
	\section{Appendix}\label{sec:appendix}
	\input{10-appendix}

	
\end{document}

%% file: 01-introduction.tex
Information security and safety has become one of the most important fields of research in computer science. Many companies and entities need information security to keep their sensitive information safe. One of the main concerns in this field is to have secure information transmission over communication networks. Today, the most common way to keep data secured is to encrypt them before transferring through the network \cite{Stallings2014security}. As a result, many protocols with different encryption methods and obfuscation techniques have been developed and implemented to improve users' privacy in networks.

Two of the most commonly used protocols to secure communication over the network are Transport Layer Security (TLS) and its now-deprecated predecessor, Secure Sockets Layer (SSL). These protocols employ cryptographic schemes to provide security for communications over networks \cite{dierks2008transport, barnes2015deprecating}. However, despite using these protocols, countermeasures have been taken into action to reveal users' identities and activities and extract other private information \cite{yen2012host}.

From an operational viewpoint, this extracted information can be used by Internet Service Providers (ISPs) to provide different levels of Quality-of-Service (QoS) \cite{roughan2004class}, and by network operators for detecting anomalies and frauds more readily \cite{kou2004survey}. However, this information leakage makes users warier while browsing the Internet. Hence, to increase the users' privacy, information leakage needs to be prevented, so it is necessary to determine the \emph{leaker parts} of data.

Despite reaching high security, many of the encryption protocols have shown to be vulnerable to cyber attacks \cite{sirinam2018deep, cai2012touching, kim2008internet, lotfollahi2017deep}. Several studies have proven that it is possible to classify users' identity and activities on the Internet from encrypted network traffic \cite{wang2017endtoend, Pacheco2019TowardsTD,lotfollahi2017deep}. Some investigations show that while network protocols are supposed to prevent information extraction, some parts of their metadata can be exploited for extracting information. This metadata contains various characteristics of data, \eg, length of packets, or initial information of a connection, like \textit{clienthello} and \textit{serverhello} and seems to be highly correlated with some flow information. For example, Bernaille \etal~performed online traffic classification of a TCP flow with only the first five packets of the connection \cite{Bernaille2006}. Moreover, Lin \etal~presented a traffic classification only based on packet size and port numbers, ignoring packet payload completely \cite{LIN2009packetsize}. Therefore, it can be concluded that there are specific parts of traffic flow from which information is more likely to be leaked. We refer to such parts of traffic as the \emph{leaker parts}.

In spite of their effectiveness, the studies mentioned above use a prior guess about which parts of a traffic flow leak the most information. However, it is more comprehensive to propose a technique to determine the leaker parts of traffic, without imposing a prior guess. By doing so, one might detect dependencies that are not known apriori. In this study, using techniques known as the \emph{interpretability} of machine learning (ML) models, we aim to find the leaker parts in a more systematic way.


More precisely, we aim to detect the leaker parts of network traffic using ML methods and deep learning (DL) techniques to extract information about the websites and domains users visit. These algorithms have taken more into account in such problems due to their incredible power of data extraction and pattern recognition on large datasets \cite{rimmer2017AutFeatureExtraction, boutaba2018comprehensive}. In fact, they need either a good prior knowledge about data or a relatively big dataset of samples. 
Because of the surge of data on the Internet and since we do not want to use prior knowledge, a sufficiently large and reliable labeled dataset with a similar distribution as the real data is required. Such a dataset can make the proposed method reliable and generalizable. However, to the best of our knowledge, no such dataset was publicly available, satisfying our traffic classification quality criteria. Hence, we try to gather such a dataset.

In this work, after collecting suitable datasets, we focus on investigating how the information is being leaked. To this end, we propose an iterative framework to find the leaker parts of the TCP traffic flow containing TLS data. The proposed iterative procedure consists of the following steps.

\begin{enumerate}[label=(\roman*)]
    \item\label{step-uno} Extract information from the data fed into learning methods by classifying input data into some given classes. The classes should be chosen based on the privacy measure we aim to investigate. For example, if we are interested in the leakage of the user's behavior from the transmitted bytes, we can consider a fingerprinting problem  with the classes chosen from the set of different webpages visited by the user.
    \item Find the leaker dimensions by applying some ML \emph{interpretability} techniques on the classification model, trained above. This interpretation leads us to the importance of each input dimension (\ie, byte) in data classification.
    \item Omit the leaker dimensions, found above, from the data. This omission can be done by masking related dimensions or any other information removal methods. Then, repeat from step \ref{step-uno}, until no significant information can be extracted.
\end{enumerate}

Since in this work, we are interested in finding information leakages related to the user's behavior, as the classification problem of Item~\ref{step-uno} above, we will conduct a \emph{web fingerprinting} (WF) attack on the user's traffic. 

Fingerprinting, in general, can be defined as detecting patterns specific to each class of data and understanding how different groups of data are distinguished from one another. In WF, this classification can be based on the user's identity \cite{yen2012host}, visited domain \cite{Rimmer2018automated}, protocols \cite{Bernaille2006}, applications \cite{lotfollahi2017deep}, or any other information related to user's identity or behavior \cite{Nikiforakis2013cookie}.

Considering the above descriptions, our contribution in this paper can be summed up as follows.
\begin{itemize}
    \item Before delving into the main focus of the paper, we have collected and labeled two separate datasets for two different settings: (1) a dataset including generated packets labeled by the domain from which the packet is loaded (referred to as the \emph{keeper domain} dataset), and (2) a dataset including generated packets labeled by the domain from which the original page is invoked (referred to as the \emph{caller domain} dataset). For example, if we browse a page from \textit{domain A} and in this page, there is some content from \textit{domain B}, the keeper domain of this content is \textit{B} and the caller domain is \textit{A}.
    
	These datasets have been collected using two different crawlers with different methods of request generation. The traffic generated by both methods are collected using \emph{namespace technology} as a lightweight mechanism for resource isolation, available in the Linux kernel \cite{namespace}. To construct our dataset, we collect raw packet-level data from 53 and 72 different domains, for datasets (1) and (2), respectively. The result was a sufficiently large dataset with a fair distribution over domains. The collection of such dataset lasted over three months.
    
    \item To the best of our knowledge, this is the first work that employs interpretation methods of machine learning models to investigate the (possible) vulnerabilities of network protocols. To this end, we propose a general framework that can be used to study almost any data transmission protocols.
    
    \item As a demonstration of the proposed framework, in this paper, we have investigated new details about how the TLS traffic leaks information. As a result, it makes researchers being aware of the leaker dimensions (\ie, bytes) and can inspire them to propose novel methods to prevent such leakage. 
    
    \item Applying the proposed method on the TLS protocol confirms that the TLS encryption method itself has some information leakage. In fact, at the end of our proposed procedure, it is revealed that the leaker dimensions not only include metadata but also include the encrypted data. Surprisingly, we have noticed that the amount of leakage of these encrypted parts is not negligible, compared to the leakage of metadata.
\end{itemize}

The rest of the paper is organized as follows. In Section~\ref{sec:relatedWorks}, we review some of the most important and recent studies on traffic classification in detail. In Section~\ref{sec:Background}, we present the essential background on machine learning used in the proposed method, including deep learning and decision tree. In Section~\ref{sec:dataset}, we discuss the process of dataset collection and its various aspects, \eg, the reason for using two different datasets. Section~\ref{sec:Methodology} presents our proposed iterative procedure and the details of employed classifiers. The results of our experiments on the gathered data are elaborated in Section~\ref{sec:results}. Finally, we conclude the paper in Section~\ref{sec:conc}, briefing what we have learned throughout this study and how to further improve it.

%% file: 02-related-works.tex
As mentioned before, network traffic classification is an important task for network operators to manage the network. Some of these tasks include QoS, service differentiation, capacity management, and performance monitoring \cite{boutaba2018comprehensive}, to mention a few. Using these classification techniques, network operators can identify unknown or suspicious network activities that are likely to be a security attack. In addition, some enterprise services may need a higher priority than other traffic. This task can also be addressed using an efficient traffic classification system.

Traffic classification can be defined as assigning a label, from a pre-defined, finite set of labels, to a particular network traffic object, \eg, a traffic flow, or a single packet. These labels may be website domains, the type of application such as Email, SFTP\footnote{Secure file transfer protocol.}, or Tor, as well as a group of services based on their QoS requirements \cite{roughan2004class}. 

A comparison of studies in the field of network traffic classification is difficult because there are no commonly accepted datasets due to different dataset gathering methods and their available features \cite{rezaei2019deep}. Also, the reported performance metrics of the studies differ significantly from each other.

Previous network traffic classification methods can be roughly categorized into five groups; namely, \emph{port-based}, \emph{packet inspection}, \emph{host behavior-based}, \emph{statistical and machine learning-based},  and \emph{cipher-based}. The rest of this section gives a brief overview of the above methods and summarizes some of the previous studies in each group.

\subsection{Port-Based}
In this method, the source and destination port numbers of a packet are used to identify its application based on IANA\footnote{Internet assigned numbers authority.} registered port numbers. The port numbers of a packet can be simply accessed from the TCP header, which is not encrypted. The main advantage of this method is that extracting the port number is not time-consuming or computationally burdensome, thereby making it a fast method compared to other methods. Despite its simplicity, due to traffic embedding and misuse of well-known port numbers to bypass traffic blocking and firewalls, nowadays, this approach is not considered as a reliable method for identifying applications. However, some studies have utilized the port number along with other techniques, \eg, see \cite{kim2008internet} and \cite{bakhshi2016internet}. 
	
\subsection{Packet Inspection}
	This method, also known as \emph{deep packet inspection (DPI)}, leverages the patterns and signatures available in the packets' application layer. They inspect deeper into the packets than just looking at structured data available in packet headers. While earlier systems employed string matching for this task, modern payload scanning tools use regular expressions (\eg, Snort, bro, trippingPoint) \cite{sherry2015blindbox}. For instance, \textit{nDPI} \cite{deri2014ndpi} is an open-source tool for traffic classification based on this method. In a comparison done by \cite{bujlow2015independent}, \textit{nDPI} and \textit{Libprotoident} achieved the best classification performance among other similar open-source softwares.
	
	It should be noted that the above tools rely on the assumption that they have access to unencrypted data. However, with the rapid growth of encryption protocols, inspection tools faced a significant challenge \cite{naylor2014cost}. To resolve such issues, these methods perform string matching on bytes sequences in packet payloads. As with Port-based methods, due to the rapid change and growth of applications and websites and also methods used for traffic encryption, DPI may fall short in maintaining the ever-growing signatures of packets \cite{erman2007identifying}. Another drawback of this method is its need for more storage and computational power compared to port-based methods. 

	
	
\subsection{Host Behavior-Based}
	In this technique, the traffic monitoring point is moved to the backbone network, a network that connects a couple of subnetworks, \eg, hubs at the physical layer, bridges at the data link layer, and routers at the network layer. Different applications have different communication patterns. For instance, clients use an identical port for connecting to a mail server, while a P2P host communicates various peers using different port numbers. Karagiannis \etal~\cite{karagiannis2005blinc}, in order to classify traffic flows, distinguished Internet applications by observing their connections' source and destination, and compared them with pre-collected application signatures. In another study, Bermolen \etal~\cite{bermolen2011abacus} used Support Vector Machine to identify P2P-TV applications by monitoring the number of packets and bytes transferred among peers over short time periods.
	
\subsection{Statistical and Machine Learning-Based}
	Generally, these methods, as the name suggests, extract statistical features from a given flow of traffic or a single packet. A flow (also called session or trace) is a set of consecutive packets communicated between two network end-points. Using features extracted from every session, the label of each one can be predicted. Machine learning methods play a pivotal role in this classification procedure. Prior to the rise of Deep Learning, early studies applied classical ML algorithms for this task. Panchenko \etal~\cite{panchenko2016website} proposed an algorithm, named CUMUL, which extracted discriminative features using the cumulative sum of packet lengths as a representation of each flow. Shen \etal~in \cite{shen2019webpage} presented a method based on cumulative lengths of first 100 packets of each flow in order to do fine-grained webpage fingerprinting using a $k$-Nearest Neighbor ($k$-NN) classifier, enabling them to achieve an accuracy of up to 91.6\%.
	
	Further development of Deep Learning methods and the advent of graphical processing units (GPU) are among the main factors paving the way for taking DL methods into account for getting more insights from the ever-increasing volume of Internet traffic data.
	Lotfollahi et. al  \cite{lotfollahi2017deep} proposed a method named \textit{Deep Packet} for both application identification and traffic characterization tasks. They implemented a $1$-dimensional convolution neural network and a stacked autoencoder on the UNB ISCX VPN-nonVPN dataset \cite{draper2016characterization} for classifying packets and reached 98\% and 93\% accuracy in application identification and traffic characterization, respectively.
	
	Intrusion and malware detection are another well-known areas in this field that aims at distinguishing malicious traffic and applications from benign ones, which is an important task in network security. Wang \etal~in \cite{wang2017malware} by using only the first 784 bytes of each flow, considering it as a $28 \times 28$ image, reached a maximum accuracy of 99.41\%. 
	Soltani \etal~proposed a framework that employs deep learning models to detect malicious traffic \cite{Soltani-DID-ArXiv20}. This framework is later extended to be able to detect zero-day attacks \cite{Soltani-adaptable-ArXiv21}. 
	
	Recently, hybrid Deep Learning models (\eg, a combination of a classic Machine Learning method and a Deep Learning-based one) are used in this field.
	For instance, Lopez \etal~in \cite{lopez2017network} employed an RNN, a CNN, and a combination of both as a hybrid model, achieving the best F1-score of 95.74\% and accuracy of 96.32\%. However, their labeling procedure is highly questionable due to using nDPI for establishing ground truth \cite{deri2014ndpi}. 
	Also, Aceto \etal~implemented a few Deep Learning architectures including SAE, 1D-CNN, 2D-CNN, LSTM, and a hybrid of LSTM and 2D-CNN, on encrypted mobile traffics \cite{aceto2019mobile}.


\subsection{Cipher-Based}
The TLS was created for obfuscating the traffic flowing through networks in order to keep information private. The encryption techniques used in the TLS make the patterns of different classes of data more indistinguishable on the ciphertext space, making the traffic classification task more difficult. However, despite expectations, since the beginning, it has always been the subject of new attacks, threatening security provided by the TLS. For instance, M{\"o}ller~\etal~devised the POODLE attack that exploited an interoperability fallback in SSL 3.0 \cite{moller2014poodle}. Aviram~\etal~utilized the Bleichenbacher adaptive chosen-ciphertext attack to crack the security of the TLS 1.2 and SSLv2 \cite{Aviram2016DROWNBT}. The ROBOT attack is another variation of Bleichenbacher attack that B\"{o}ck~\etal~have introduced \cite{10.5555/3277203.3277265}. Using ROBOT, an attacker can sign any desired message on behalf of the victim, while s/he does not have any additional access to the private key of the victim. They have reported that one-third of the top 100 domains in the Alexa are vulnerable to ROBOT attack. After reporting such a flaw, the RSA PKCS \#1 v1.5 standard has been deprecated.




The term TLS fingerprinting has been widely used in the academic literature \cite{Frolov2019TheUO, Holz_2016, 10.1109/ARES.2015.35, 10.1186/s13635-016-0030-7, 10.1145/3278532.3278568, tls-android, 10.1145/3355369.3355601}. Recently, Anderson~\etal~gathered a knowledge base for TLS fingerprinting \cite{10.1145/3355369.3355601}. Their dataset not only includes tunnelled traffics of web applications fingerprints, but also they put a step forth and extend the dataset by the data collected from a different set of applications, \eg, storage, communication, system, and email.

Although Deep Learning has attracted researchers for traffic classification, there have not been many studies on traffic classification using the ciphertext of the TLS. The TLS connections have a set of specifications by design. So far, only some of the plain features such as those in TLS extensions are being used for traffic classification tasks. However, there are other features, including hidden ones, in the cipher, \eg, the chosen cipher suite in the connection, the chosen initial vector (IV), encrypted payload, etc., that are suspected to be valuable for the classification task. To the best of our knowledge, this is the first work in which such information is investigated besides the plain features of the TLS in traffic classification. We call this method \emph{cipher-based} traffic classification.



%% file: 03-machine-learning-methods.tex
In this study, we employ ML techniques to extract information from raw data, which is network traffic in our case. In particular, we are aiming at information that is not easily recognizable by humans. Then, we employ interpretation techniques studied in the ML literature, to find the dimensions in data that are more likely to leak information. This whole process enables us to recognize the vulnerability point of network protocols. In the following, we introduce the ML methods we will employ to investigate the TLS protocol.


One of the ML methods used in this work is the \textit{Decision Tree}. The output of this method is a tree of ``yes-no'' questions, called binary questions. After answering each question, the tree leads us to either the final analysis (\ie, a data class) or asking another question, to finally output the desired analysis \cite{quinlan1986induction}. There are some methods to prepare the resulting question tree based on training data. These methods try to minimize the average number of questions asked for determining the output (optimal analysis) while providing an analysis that is as close as possible to the desired output. Using these methods, the questions are asked in a way that after each question, the entropy of the two resulting distributions is minimized. Finally, using the test data, the accuracy of the resulting tree can be examined \cite{hu2011rank}.

One advantage of using the decision tree over other ML methods is that it can be easily interpreted despite its remarkable ability to extract information. In fact, given the decision tree of the questions, and in particular, considering the most primitive questions, one can determine which dimensions of input data have the most impact on determining the output. Thus, in this way, we can readily identify which dimensions contain more information about the output label \cite{hastie2009elements}.

Another ML method used in this study is deep learning, which is based on artificial neural networks. An artificial neural network is a network of computational nodes, called \emph{neurons}, that are interconnected according to some architecture. Each of these neurons receives the output of some other neurons as inputs, computes a weighted sum of them, and then applies an activation function (\eg, Sigmoid, Tanh, or ReLU), and outputs the result \cite{schmidhuber2015deep}.

The architecture of multilayer neural networks is in a way that the neurons are usually partitioned into sequential clusters, called \emph{layers}, in which the inputs of the neurons of each layer are the outputs of the neurons from the previous layer. First layer neurons use input data as their input and neurons of the last layer generate the network's output and present the analysis. Using the training data and their desired output analysis, employing the backward propagation algorithm, and an optimization method, the network parameters (weights of neurons) are determined. Eventually, the performance of the network is evaluated using the test data.

As the number of layers in these networks grows, their training procedure faces particular challenges. Such networks are called deep neural networks, and their learning process is called deep learning. Deep neural networks learn to present a more abstract representation of the input data in each layer than the previous layer. In this vision, after training a deep network, the raw input data itself represents the data in the least abstract way, and the output of the network, trying to be the desired analysis, is the most abstract representation of the input data in the network \cite{Goodfellow-et-al-2016}.

As the number of neural network layers increases, the whole network's overall function becomes capable of approximating a vast range of functions, given appropriate parameters. Deep learning has performed well in many applications and has been able to extract complex patterns and relations hidden in the input data \cite{Goodfellow-et-al-2016}.

For different types of data, different classes of neural network architectures have been presented. One of these classes, called the convolutional neural network (CNN), is able to reduce the computational complexity significantly. These neural networks are commonly used for stationary data types such as image and sound. In these neural networks, the weights between the layers are very sparse, and the neurons of each layer are affected by only a small number of locally nearby neurons from the previous layer. Furthermore, non-zero weights are replicated, a technique called parameter sharing in the neural network literature. As a result, in each layer, the same filters are applied to different parts of data to obtain the next layer, which leads to lower computational complexity. These networks have shown an outstanding performance on stationary data \cite{kim2014convolutional}.

Another class of neural network architectures, called recurrent neural network (RNN), is mainly used to analyze sequential data. In an RNN, data elements are given to the network one by one, and the network produces output for each of them. Their difference with conventional networks is that in some layers, the inputs of neurons are the outputs of the precursor neurons in addition to the previous value of neurons in the same layer. So, these neurons maintain a history of inputs. Thus, these networks can also use inter-elemental dependencies to provide appropriate analysis. In these networks, when there is a need to calculate one output for the entire input data, the last output is considered the only output for both training and evaluation \cite{graves2012supervised}.

Because of the inability of regular neurons to detect the long-term interdependencies between elements, recursive neural networks typically use long short-term memory (LSTM) units instead of regular neurons. With their novel structure, these neural networks proved to be able to exploit the long-term dependencies between the input data elements in the analysis \cite{graves2012long}.

%% file: 04-dataset.tex




Since we aim to examine two different settings, our dataset needs to accompany our goal. Furthermore, ML methods work ideally with large datasets \cite{boutaba2018comprehensive}. To this end, with the help of Scrapy \cite{Scrapy} and Selenium \cite{Selenium}, we have developed a crawler to gather two datasets, based on each setting's needs. What follows is the specifications of each setting and their dataset statistics.


\subsection{Keeper Domain}
\label{subsec:sess-des}

In order to analyze the data protection inside the TLS channel, we propose a fingerprinting attack with the purpose of examining the destination of traffic session based on its encrypted payload, with or without considering the TLS headers.
 
The dataset for this attack is generated with the \emph{Scrapy} tool. The destinations are pages from websites shown in Fig.~\ref{fig:scrapy}. However, we keep the crawler away from requesting inner objects of the initially requested page, making the dataset simpler. We refer to this dataset as the \emph{keeper domain} dataset since we keep the crawler away from calling inner requests of pages we are visiting. We captured the traffic generated by the crawler with the \textit{tcpdump} tool. Finally, the traffic of each session is saved into a single pcap file.



As discussed by Zliobaite \etal~in \cite{zliobaite2016}, data and the way it is encrypted, processed, and exchanged in networks are subject to alter over time. These changes affect the efficiency of machine learning methods. This phenomenon, called \emph{concept drift}, is well known in the ML literature \cite{Bernaille2006}. In order to prevent ML models from being affected by this alteration, data should be collected and fed to the model over a long enough time period. Hence, to mitigate the effect of traffic concept drift, our datasets were captured from target websites over the course of three months.




\subsection{Caller Domain}
\label{subsec:usr-bhvr}
Our second goal is to detect which website the user visits, from the adversary's point of view, who is sniffing the whole traffic flow. In each webpage, there may exist additional references in the HTML Document Object Model (DOM) which leads the browser to send more requests. When a user visits a webpage, additional parts of DOM, such as images and javascript, are also fetched. So the traffic flow is intertwined with sessions other than the initially requested session. To this end, the Selenium crawler has been used which sends all requests referenced in DOM just as a real user when surfing a webpage, so it generates a more realistic traffic flow than Scrapy. 

While generating traffic with the Selenium crawler, because of the additional information being collected, we need a way to fully isolate every Selenium process in order to label each session with its corresponding website name. For example, consider the case that we are trying to crawl ``youtube.com,'' and the browser establishes a session to ``google.com'' for loading the youtube page. The goal is to label both sessions with ``youtube.com''. As a solution, we instantiate a new chrome process for crawling every domain. This way, we only need to isolate each of these chrome processes' network traffic to reach our goal. We refer to this dataset as the \emph{caller domain} dataset since we call inner requests of a specific page in addition to the initial request of that page and collect their traffic as well, in contrast to the keeper domain dataset where we collect only the initial requests' traffic of crawled webpages.

\begin{figure}[h]
	\includegraphics[width=0.9\textwidth]{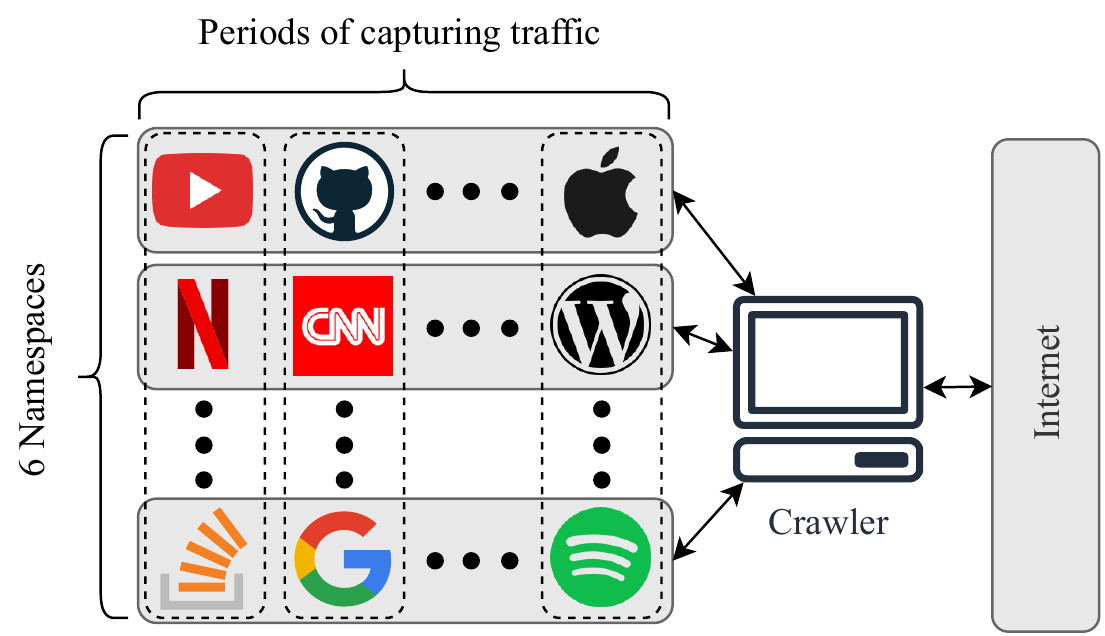}
	\caption{Six parallel instances of Selenium crawler choose a website every time, crawl 100 pages from each, capture their traffic in the respective namespace, and label them accordingly. The crawlers repeat this process until the whole data is collected.}
	\label{fig:namespace}
\end{figure}

For capturing network traffic of the desired process in the operating system, we employed \textit{network namespace} or \textit{netspace} technology. As illustrated in Fig.~\ref{fig:namespace}, for every process that runs at the same time, we assign a unique virtual interface. This way, we can label every transferred packet through a virtual interface with its corresponding label.

For generating traffic, we have used the Chrome browser's driver. The final crawler ran on two systems with six threads on each; every thread crawled a domain with 100 random pages in a unique namespace and collected their traffic every time. On average, there are about 2000 pages crawled from every domain. Also, there is a 6-second delay between collecting every page. In this way, the destination server does not recognize the crawler as a bot. Actions related to considering traffic concept drift have also been taken in this setting. The traffic generation and collection process with this tool took about three months to complete.

\subsection{Datasets' Statistics}
\label{subsec:properties}

As mentioned before, for both keeper domain and caller domain datasets, we use the domain of the requested page as the label of all the generated traffic sessions. In the keeper domain dataset, the crawler does not follow the links that exist in the requested page. In contrast, the caller domain dataset is captured by letting the crawler follow the internal links of the requested page. In this way, the latter dataset is more resemblant to the traffic generated by a user while browsing the Internet.

\begin{figure}[htp]
	\subfloat[]{

		\includegraphics[width=\textwidth]{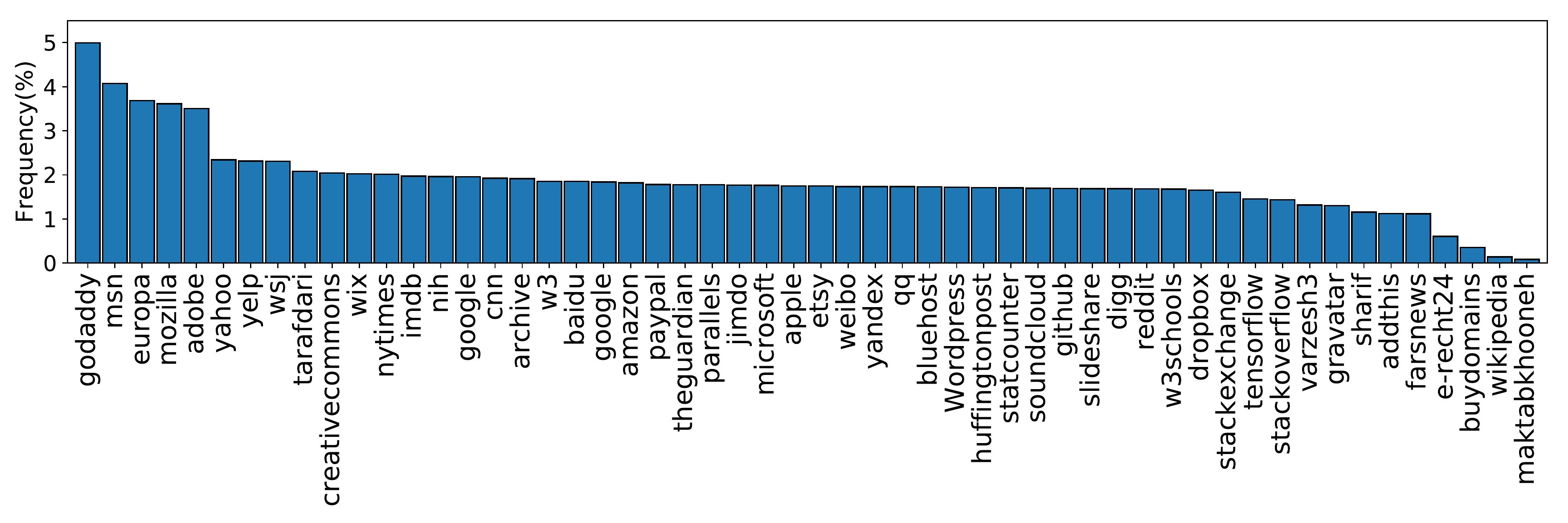}
		\label{fig:scrapy}
	}

	\subfloat[]{

		\includegraphics[width=\textwidth]{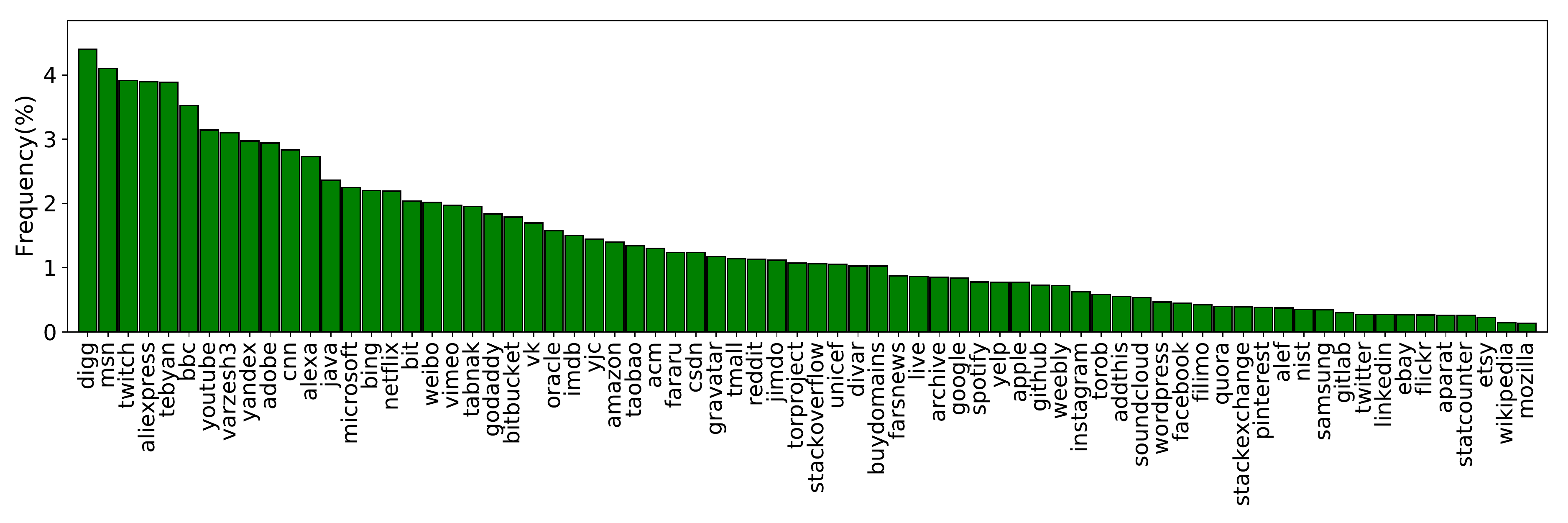}
		\label{fig:selenium}
	}
	
	\caption{Frequency of domains' sessions collected with (a) the Scrapy crawler and (b) the Selenium crawler. This figure has many details observable by zooming, and hence, it can be fully observable only in the digital version of the paper.}
	
\end{figure}

The properties of these datasets are summarized in Table~\ref{tab:prop}. In addition, the proportions of each domain's data in both keeper domain and caller domain datasets are illustrated in Figs. \ref{fig:scrapy} and \ref{fig:selenium}, respectively. Length of a particular session is defined as the number of packets it contains.


\begin{table*}[t]
	\ra{1.3}
	\begin{tabular}{|c|c|c|c|c|}
		\hline
		\textbf{Label} & \textbf{Number of Sessions} & \textbf{Number of Packets} & \textbf{Avg. Length of Sessions} & \textbf{Number of Classes} \\ \hline
		Keeper Domain              &  525,066        & 46,467,876         &     89                   &     53    \\ \hline
		Caller Domain     &    358,865      &     107,133,709    &         299               &      72   \\ \hline
	\end{tabular}
	\caption{The summary of our dataset properties.}
	\label{tab:prop}
\end{table*}
%
%

%% file: 05-methodology.tex
As mentioned in the previous sections, we aim to determine which parts of a TCP packet containing TLS data can be used to extract some information about the user's behaviour. We refer to the packets' recognizability from their encrypted data bytes as the leakage of information and the parts of raw data from which this recognition can be done as leaker parts. In this section, we are going to propose a general framework for detecting leaker parts of network protocols. Using this framework, we will be able to detect the leaker parts of TCP packets conveying TLS data. The insight obtained from our proposed framework helps researchers and protocol designers to focus on the parts of raw data that need protection the most.

Our proposed framework for detecting leaker bytes of transported encrypted packets is an iterative process described in Framework~\ref{fw:ProposedFramework}. We start Framework~\ref{fw:ProposedFramework} by applying some ML methods on the raw input data, as stated in Line~\ref{fw:line:extract_info}. For example, if we aim to investigate the leakage of a user's information to the encrypted TLS bytes, we can employ a webpage fingerprinting attack on the user's traffic. This is what we will apply in order to find the leaker dimension of the TLS protocol. Then, in Line~\ref{fw:line:Interpretability}, using an interpreting technique, applied to the ML model trained in Line~\ref{fw:line:extract_info}, we can recognize the most significant leaker bytes of the protocol. Next, we remove the most significant leaker bytes (\eg, by deleting or masking them), and repeat the above process to examine the leakage of the remaining bytes.

\begin{algorithm}
	\caption{Finding the leaker dimensions of a network protocol.}
	\label{fw:ProposedFramework}
	\begin{algorithmic}[1]
		\REPEAT
			\STATE Extract the desired information (\eg, the webpage a user visits) from the raw data (generated by a network protocol, \eg, TLS) by applying some machine learning method(s). \label{fw:line:extract_info}
			\STATE Apply an interpretability technique to ML model(s) trained in Line~\ref{fw:line:extract_info}. \label{fw:line:Interpretability}
			\STATE Using the results obtained in Line~\ref{fw:line:Interpretability}, identify the input features (\ie, here input bytes) that have high influence on the output of the trained ML.
			\STATE Remove the effect of these influential bytes (\eg, by deleting or masking them).
		\UNTIL{The performance of the ML method, trained in Line~\ref{fw:line:extract_info}, on the desired task drops significantly and no input has a significantly more impact on the ML output than others.}
		\RETURN{The whole set of influential input bytes found above.}
	\end{algorithmic}
\end{algorithm}

As demonstrated in Framework~\ref{fw:ProposedFramework}, we repeat the procedure (Line~2 to Line~5) until no bytes have a significantly more impact on information extraction than others, and the information leakage is distributed evenly across the remaining bytes. At this point, if information leakage is negligible, we can obtain data that is sufficiently secure by protecting bytes that have already been identified as leaker bytes.

In this study, we use neural networks and decision trees as ML methods to apply the above framework. The decision tree is used because it can be interpreted and so we can find leaker bytes readily. Neural networks can also be used to verify the accuracy obtained by the decision trees because of their high capability to extract complicated information. If the decision tree extracts information with an accuracy close to the neural networks, its extraction and so its recognition of leaker bytes is reliable. It is expected that when no bytes have a significantly more important role in leakage, the decision tree will not be able to reach the accuracy of neural networks. In this case, of course, we have reached the end of the proposed method.

The data used in this study, introduced in Section~\ref{sec:dataset}, are TCP segments, containing TLS packets, captured over the network. In order to reveal the leaker part of TLS packets, we propose a fingerprinting attack on the user-requested domain by observing the encrypted TLS traffic. Moreover, we would like to emphasize that the TCP header highly depends on the various factors of the crawling method. So, investigating its leakage is not useful since it is not consistent with real-world settings and prevents the model and results from generalization. Therefore, to apply our proposed method, we only examine the leakage of TCP payload.

\begin{figure}[h]
	\centering
	\includegraphics[width=\textwidth]{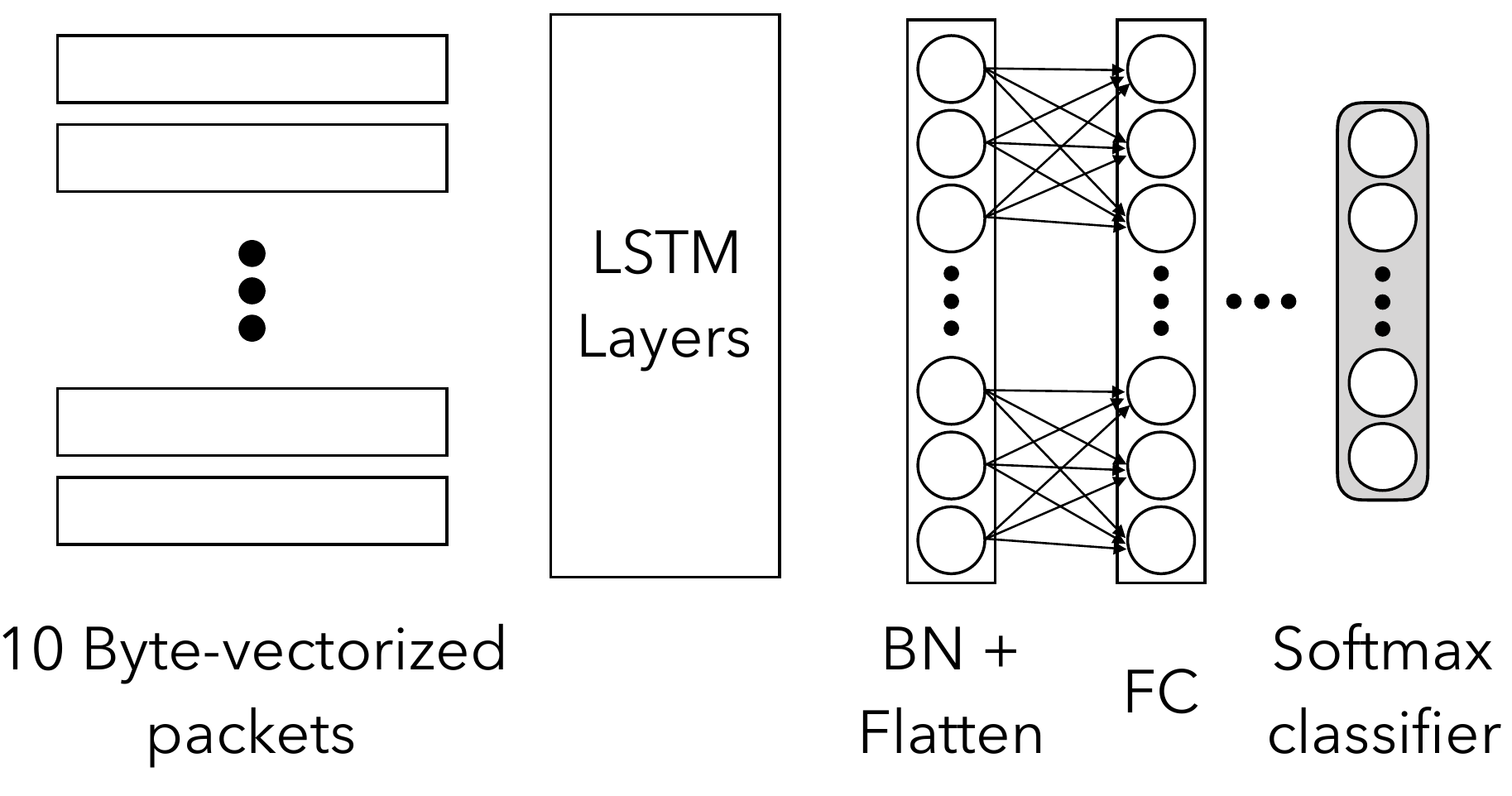}
	\caption{The general architecture of the proposed LSTM network to implement website fingerprinting attack.}
	\label{fig:lstm}
\end{figure}
\begin{figure}[h]
	\centering
	\includegraphics[width=\textwidth]{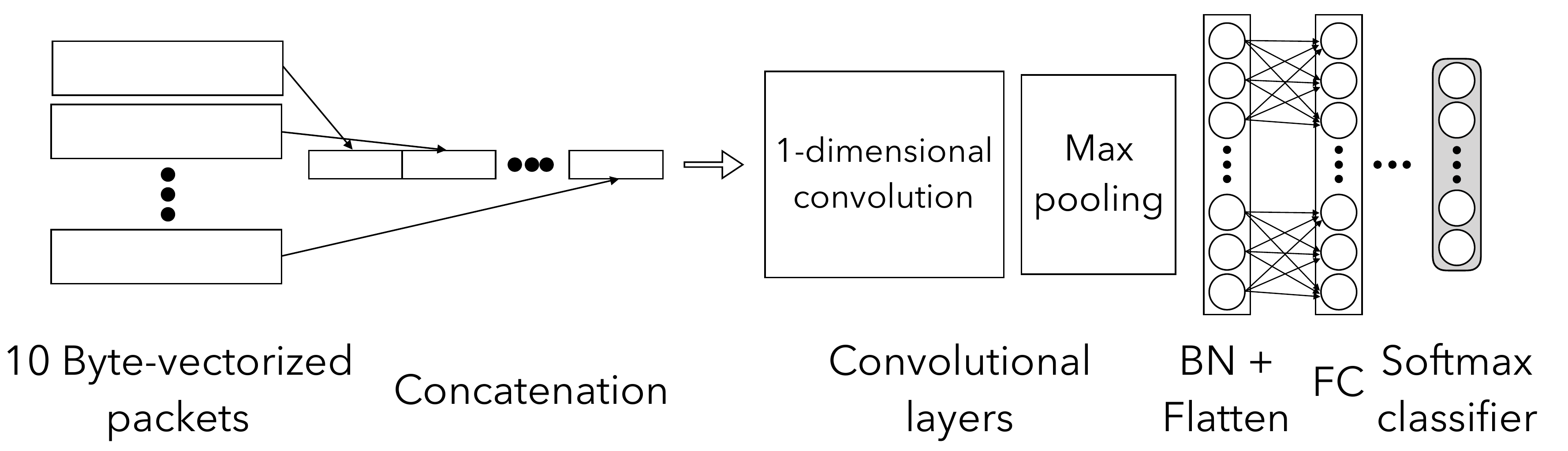}
	\caption{The general architecture of the proposed 1-dimensional convolutional network to conduct a webpage fingerprinting attack.}
	\label{fig:conv}
\end{figure}

As mentioned before, in addition to decision trees, we also use deep neural networks as an ML method to investigate TLS protocol leaker bytes. We implemented two different types of deep neural networks, as stated in the following.
\begin{itemize}
	\item
		First, we employ a network having a sequence of recurrent layers with LSTM nodes, followed by a sequence of fully-connected layers. We use a batch normalization layer between the last recurrent and first fully connected layer, which results in speeding up the learning procedure \cite{ioffe2015batch}. For regularizing the model and preventing overfitting, a dropout layer is used after the last fully-connected layer. An illustration of the network architecture is represented in Figure~\ref{fig:lstm}.
	\item
		Second, we implement a network having a sequence of 1-dimensional convolutional layers with max-pooling output followed by a sequence of fully-connected layers. We use batch normalization and dropout layers in this model to speed up the learning procedure and to regularize. An illustration of this network architecture is represented in Figure~\ref{fig:conv}.
\end{itemize}


One of the crucial factors that significantly affect the final performance of an ML method is hyperparameter tuning. A model with a poor hyperparameter set may not even converge in the training phase, while the same model with a good set of hyperparameters can lead to solid results. Hyperparameter tuning includes choosing each hyperparameter (\eg, type of the activation function, or learning rate) from a given space. The straightforward and naive solution is a brute-force search on the given space. However, this is not always feasible in practice due to the enormous resources this approach needs, and the long time it takes. So a more intelligent search algorithm is of interest. In this work, we use the Tree-structured Parzen Estimator (TPE) as a suitable algorithm to tune the hyperparameters \cite{bergstra2011algorithms}. Using this algorithm, we choose the hyperparameters of the proposed recurrent neural network and convolutional neural network from the search spaces described respectively in Tables \ref{tab:lstm-hyper} and \ref{tab:conv-hyper}.

\begin{table}[H]\centering
	\ra{1.3}
	\begin{tabular}{|c|c|}
		\hline
		\textbf{Parameter}    & \textbf{Search Space}      \\ \hline
		Optimizer             & {[}sgd, adam, adadelta{]}  \\ \hline
		Learning Rate         & {[}0.001, 0.0001{]}        \\ \hline
		\# LSTM Layers & {[}1, 2, 3{]}              \\ \hline
		\# LSTM Neurons          & {[}16, 32, 64{]}           \\ \hline
		\# FC Layers   & {[}1, 2, 3{]}              \\ \hline
		\# FC Neurons            & {[}128, 256, 512{]}        \\ \hline
		Activation Function            & {[}relu, tanh, softplus{]} \\ \hline
		Dropout Probability               & uniform(0, 0.5)            \\ \hline
	\end{tabular}
	\caption{The hyperparameter search space for the proposed LSTM network of Fig.~\ref{fig:lstm}.}
	\label{tab:lstm-hyper}
\end{table}

\begin{table}[H]\centering
	\ra{1.3}
	\begin{tabular}{|c|c|}
		\hline
		\textbf{Parameter}    & \textbf{Search Space}      \\ \hline
		Optimizer             & {[}sgd, adam, adadelta{]}  \\ \hline
		Learning Rate         & {[}0.001, 0.0001{]}        \\ \hline
		Strides               & {[}1, 4, 8{]}              \\ \hline
		\# Convolution Layers & {[}1, 2, 3{]}              \\ \hline
		\# Filter                & {[}64, 128, 256{]}         \\ \hline
		Kernel                & {[}8, 32, 128{]}           \\ \hline
		\# FC Layers   & {[}1, 2, 3{]}        \\ \hline
		\# FC Neurons            & {[}128, 256, 512{]}        \\ \hline
		Activation Function            & {[}relu, tanh, softplus{]} \\ \hline
		Dropout Probability               & uniform(0, 0.5)            \\ \hline
	\end{tabular}
	\caption{The hyperparameter search space for the convolutional network of Fig.~\ref{fig:conv}.}
	\label{tab:conv-hyper}
\end{table}

For the decision tree model, the only hyperparameter used in this study is the depth of the tree. Moreover, the training phase of the decision tree is not a very time-consuming process. So, finding the best hyperparameter of the decision tree model can be done easily by brute-force searching.

%% file: 06-experiment.tex
In this section, we aim to systematically investigate the leaker bytes of the TLS protocol using Framework~\ref{fw:ProposedFramework}, presented in Section~\ref{sec:Methodology}. We apply the proposed framework on the data captured by the methods described in Section~\ref{sec:dataset}. Our dataset consists of samples where each one is the raw bytes of the whole TCP session containing TLS encrypted data.

As mentioned in Sections~\ref{sec:dataset} and \ref{sec:Methodology}, each traffic sample is labeled with (a) its caller domain, and (b) its keeper domain. We study the information leakage of the TLS protocol using data labeled by these two methods. Each data sample contains the whole traffic of a session. Since such a data element is relatively huge, it needs to be preprocessed before usage, by the following considerations.

\begin{enumerate}
\item
In real-time systems, it is highly crucial to make a decision about the input data in a short time---for instance, DDoS\footnote{Distributed denial of service.} detection systems have to identify and block malicious network traffic as quickly as possible. Therefore, we need to extract the necessary information before the whole network traffic is transferred. So, we need to focus only on the initial packets.
\item
The data related to each sample has a relatively huge volume and contains too much redundancy. The hugeness of data can confront us with the curse of dimensionality \cite{verleysen2005curse}. Due to the high redundancy of data, in our analysis, we can only focus on some specific parts of data without significant loss in the accuracy of the result.
\end{enumerate}

\begin{figure}[htp]
        \subfloat[]{%
        \includegraphics[width=1\textwidth]{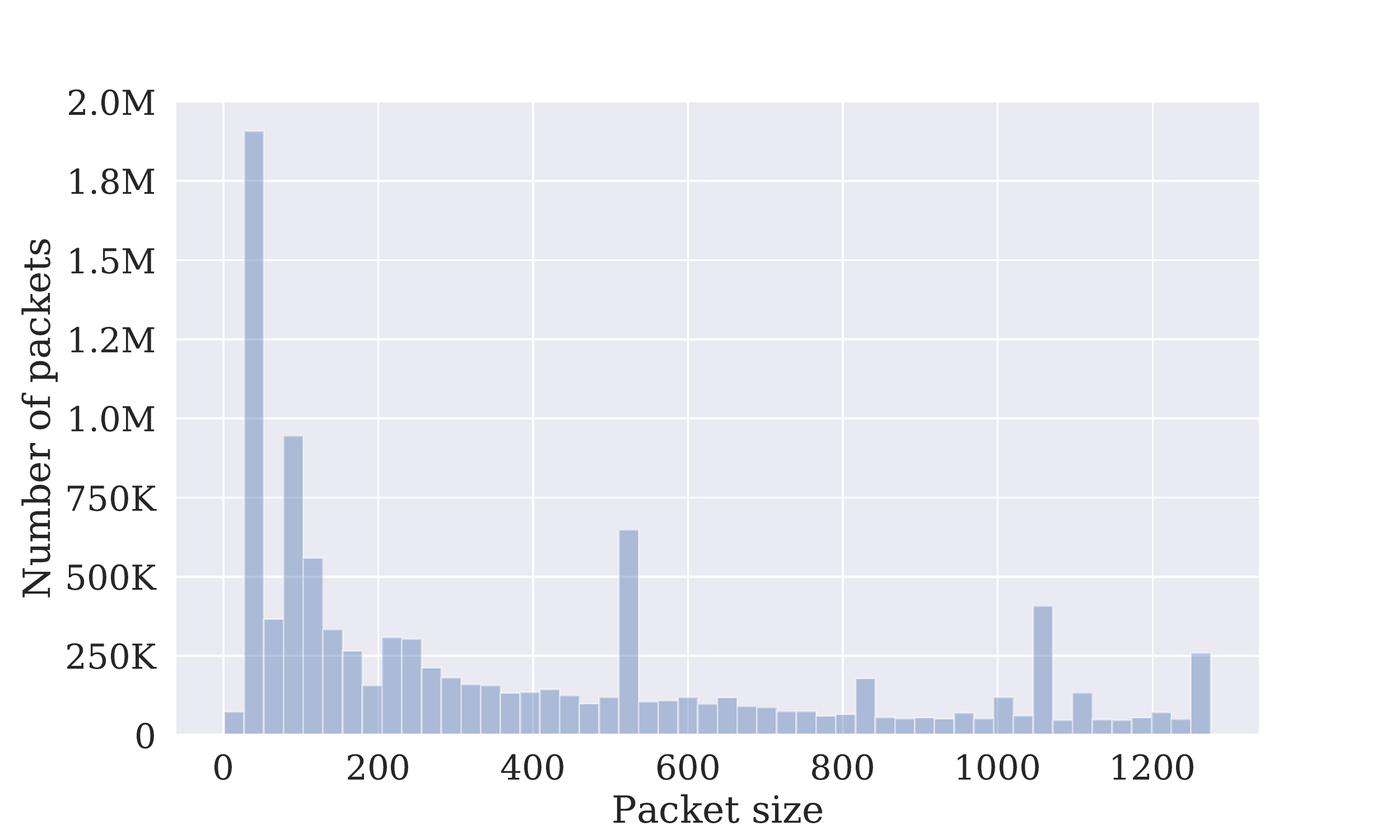}
        \label{fig:lengths1} 
    }

    \subfloat[] {
        \includegraphics[width=1\textwidth]{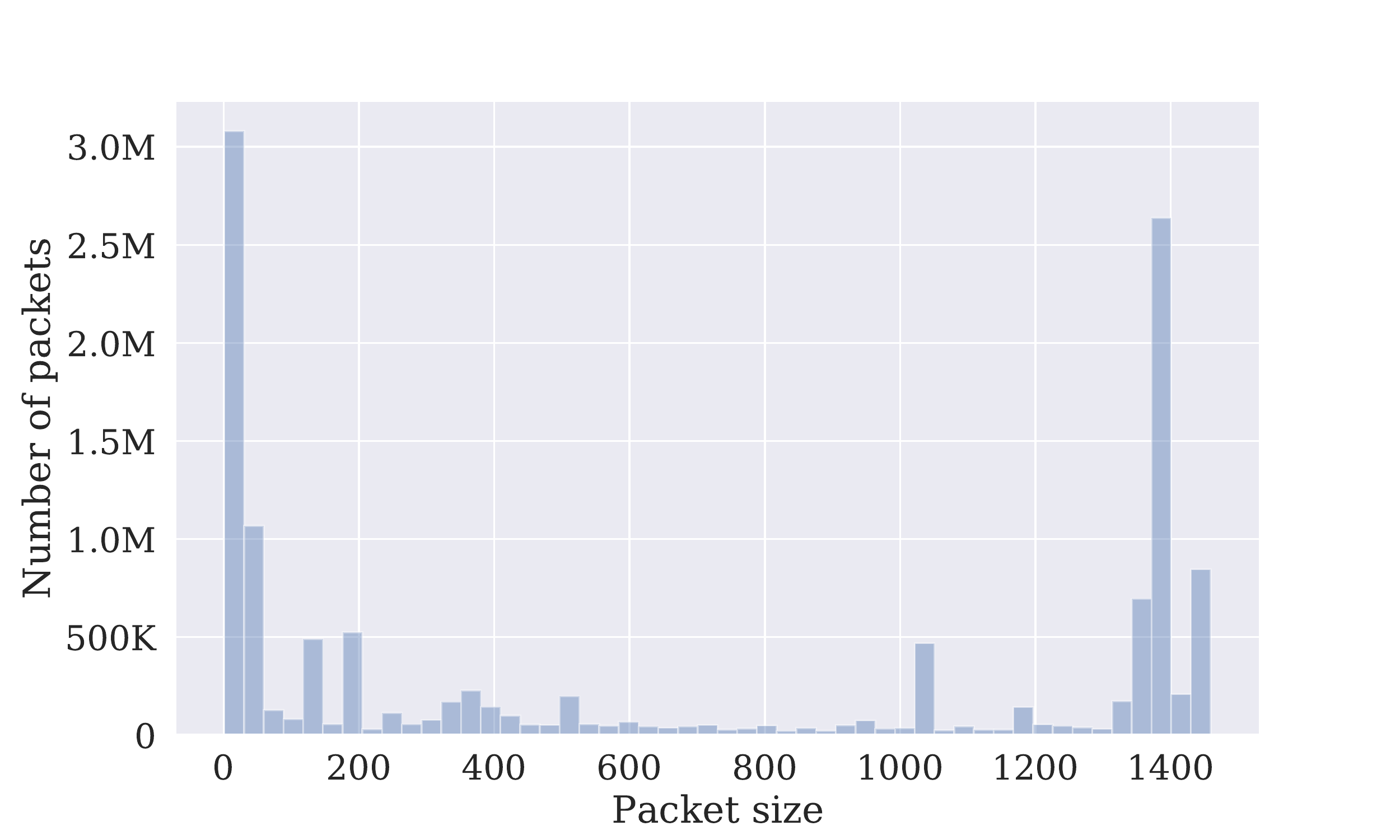}
        \label{fig:lengths2} 
    }
       
    \caption{The histogram of packet lengths in the captured datasets. Empty packets and packets larger than 95th percentile are removed from the graph.  (a) Dataset crawled by Selenium, and (b) Dataset crawled by Scrapy.}
\end{figure}

To consider the above points, we take the first 200 bytes of each session's first ten packets. Our experiments revealed that for the captured dataset, the information existing in the rest of the packets or the remaining bytes of each packet is negligible. Also, considering the histogram of packet length in our dataset, depicted in Figs. \ref{fig:lengths1} and \ref{fig:lengths2}, one can observe that the majority of packets are smaller than 200 bytes. In particular, this fact matches better to the traffic captured by the Selenium crawler, which produces traffic that we know is more resemblant to real-world traffic. Moreover, since most ML methods need fixed-length input, we zero-padded packets with length fewer than 200 bytes to make them of length 200 bytes. Finally, it should be noted that the first three segments of every TCP session belong to the handshake of TCP protocol, and since their payloads are empty, we ignore them and apply the proposed framework from the fourth segment.

As mentioned before, we aim to apply our proposed iterative framework on the captured data. So, we need to determine the initial data on which we start the algorithm, the ML models for extracting information, and the interpretation methods. In the rest of this section, we will go through the Framework~\ref{fw:ProposedFramework}, iteration by iteration.

We use three types of machine learning models for extracting information from the TLS traffic. The ML models include: \textsf{(i)} a decision tree, \textsf{(ii)} a recurrent neural network with LSTM units, and \textsf{(iii)} a 1-D convolutional neural network. Deep networks (\eg, RNNs and CNNs) are highly capable of extracting information and relations while the decision tree is not as good on the data with complicated patterns. However, the decision trees are easily \emph{interpretable}, while deep networks' interpretability is still under investigation. So, if the decision tree's extraction accuracy gets close to that of deep networks, we can assume that one can effectively find the leaker bytes of traffic by interpreting the decision tree. As a result, in the case that the accuracies of these techniques are close enough, we choose the decision tree's \emph{impact vector} as the interpretation method used in our framework.

In the following, we will describe the results of iterations and the transformations between iterations. We aggregate all the results of iterations in Table~\ref{tab:accs}. For applying the above ML models, we need to tune their hyperparameters properly. These values were chosen using the method described in Section~\ref{sec:Methodology}. The final value of the hyperparameters is presented in the Appendix. After training the ML models for the information extraction tasks (which in our case is a webpage fingerprinting task), we achieved the accuracies given in Table~\ref{tab:accs}. As we observe, the decision tree has almost achieved the same accuracy as of deep neural networks in all iterations. So, its accuracy is reliable, and we can use its impact vectors reliably as our interpretation method.


\textbf{First iteration: TCP payloads.}
In the first iteration, as discussed earlier, we start with the zero-padded first 200 bytes of packets 4 to 13 in each TCP session. We apply the ML methods introduced in Section~\ref{sec:Methodology} on such input data. The achieved accuracies of the three ML models are reported in Table~\ref{tab:accs}. It can be inferred from the results that the decision tree has achieved accuracy close enough to the accuracy of deep neural networks. So, its accuracy is reliable to be used as an interpretation method. 

\begin{figure*}[htp]

    \caption{The impact of each byte in the classification of the keeper (left) and caller (right) domain of sessions on different data samples.}
    \label{fig:imps}
	\begin{tabular} {c c}
		\subfloat[TCP Payloads]{%
			\includegraphics[width=0.45\textwidth]{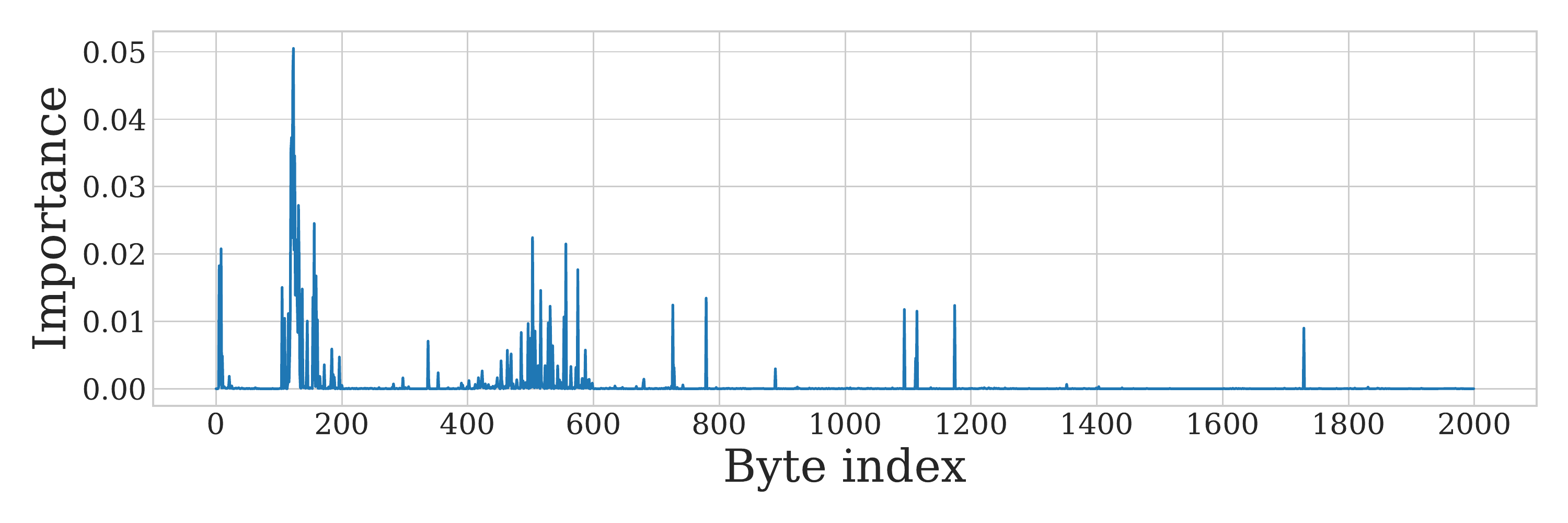}
			\label{fig:keeperImp1}
		}
		&
		\subfloat[TCP Payloads]{%
			\includegraphics[width=0.45\textwidth]{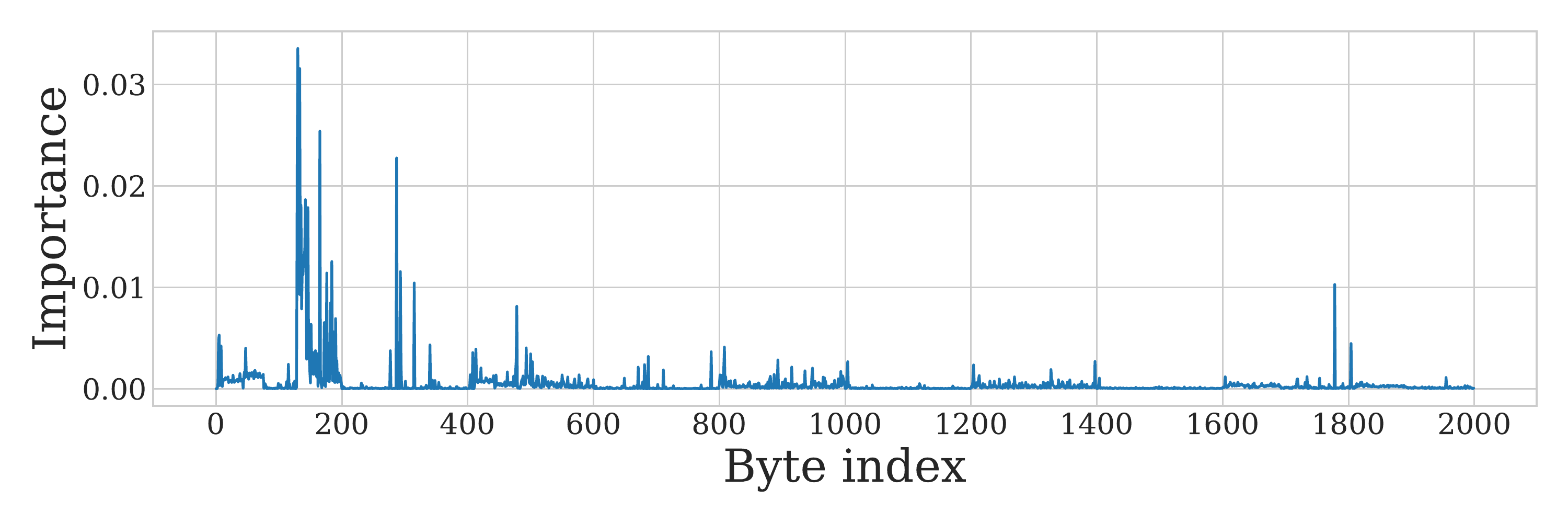}
			\label{fig:callerImp1}
		}
		\\
		\subfloat[The TLS Packets] {
			\includegraphics[width=0.45\textwidth]{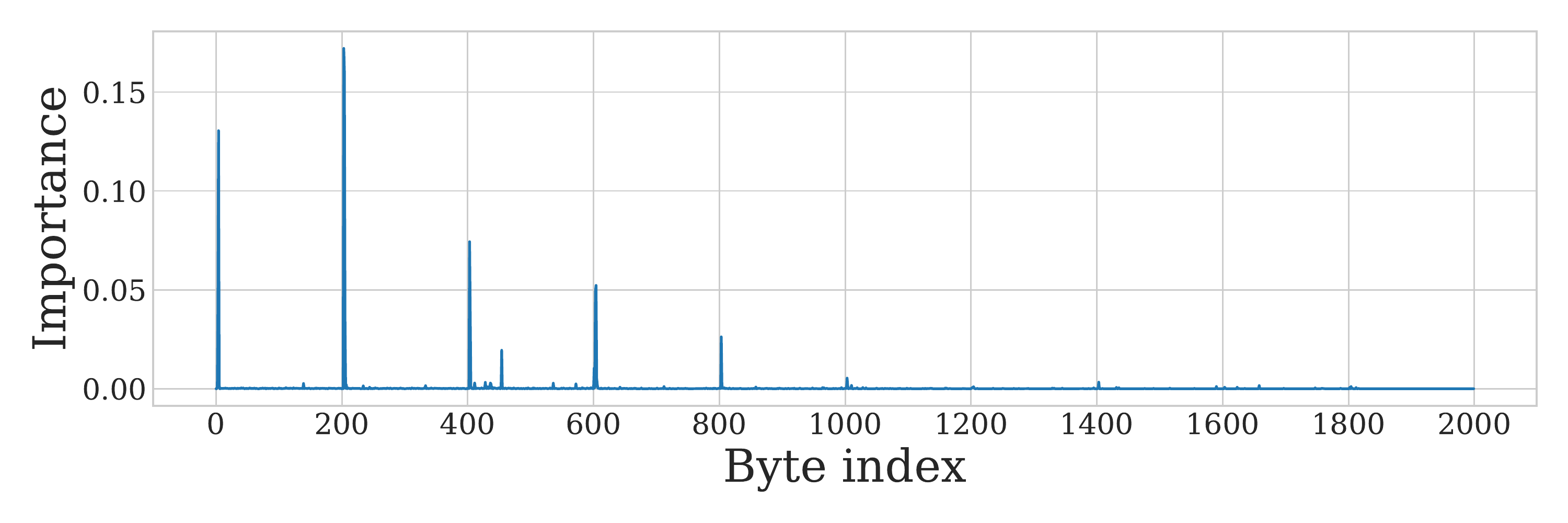}
			\label{fig:keeperImp2}
		}
		&
		\subfloat[The TLS Packets] {
			\includegraphics[width=0.45\textwidth]{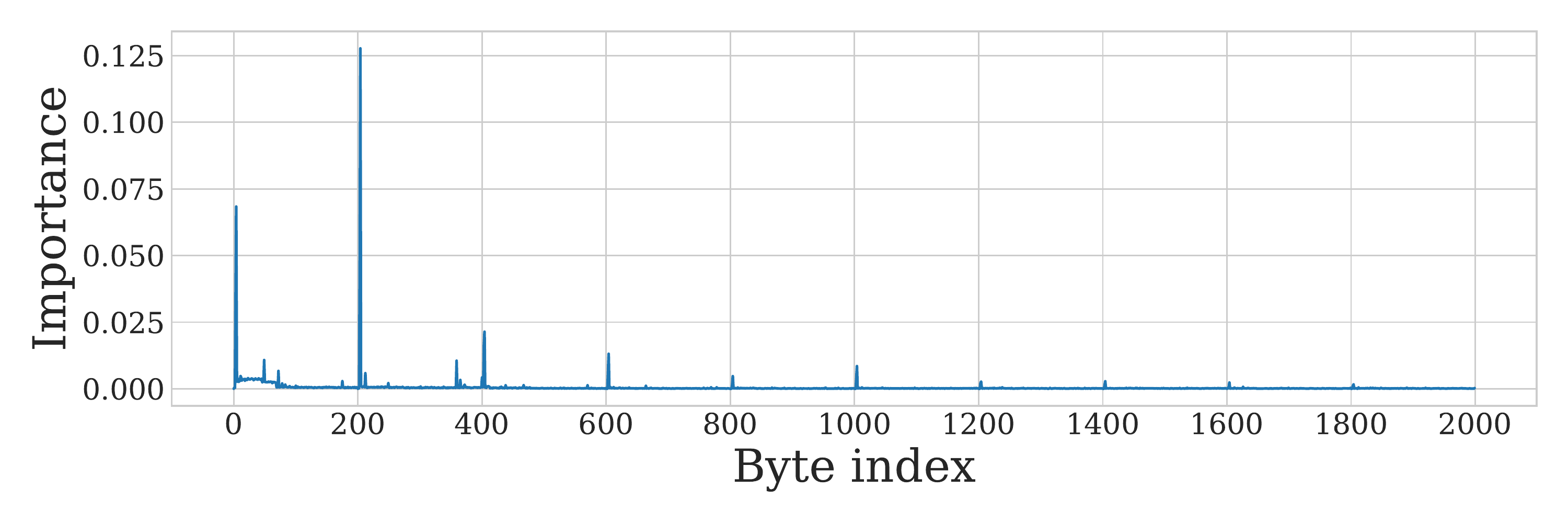}
			\label{fig:callerImp2}
		}
		\\
		\subfloat[The TLS Payloads]{%
			\includegraphics[width=0.45\textwidth]{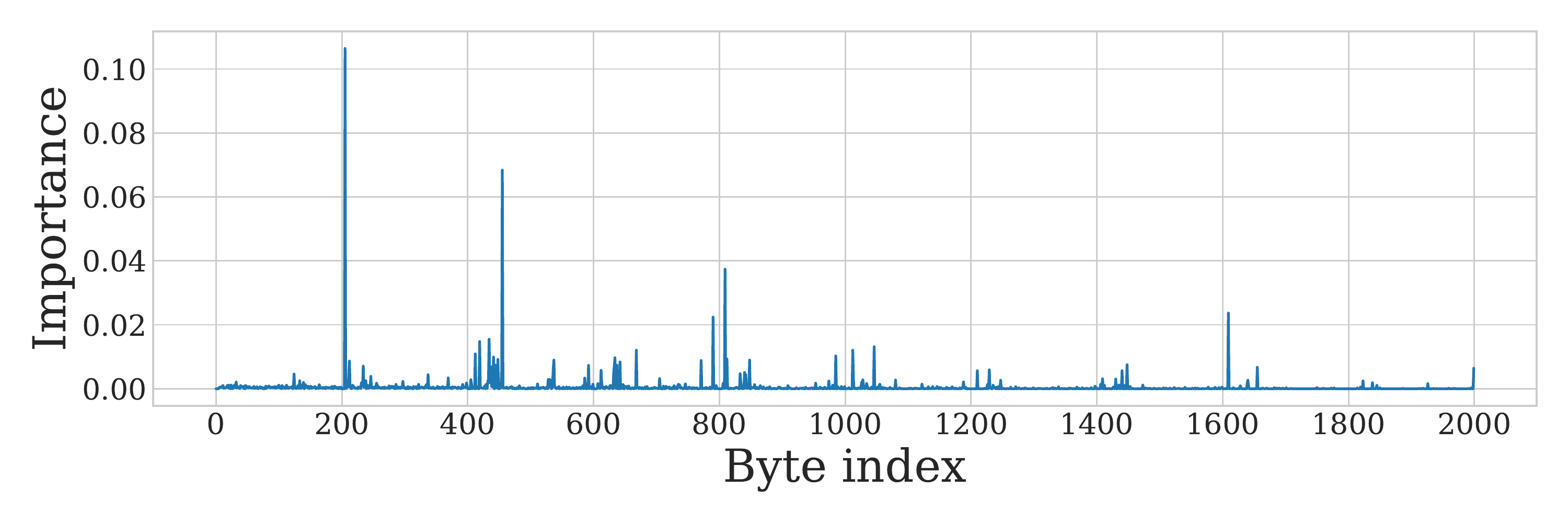}
			\label{fig:keeperImp3}
		}
		&
		\subfloat[The TLS Payloads]{%
			\includegraphics[width=0.45\textwidth]{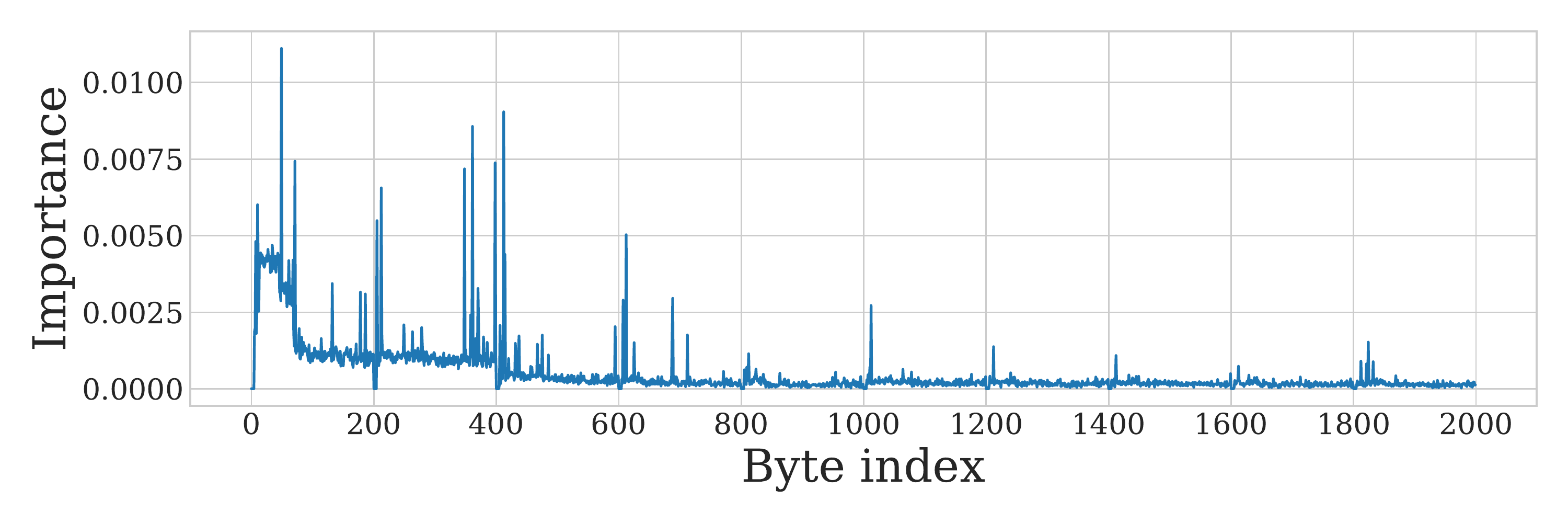}
			\label{fig:callerImp3}
		}
		\\
		\subfloat[The TLS Payloads with IV vectors masked]{%
			\includegraphics[width=0.45\textwidth]{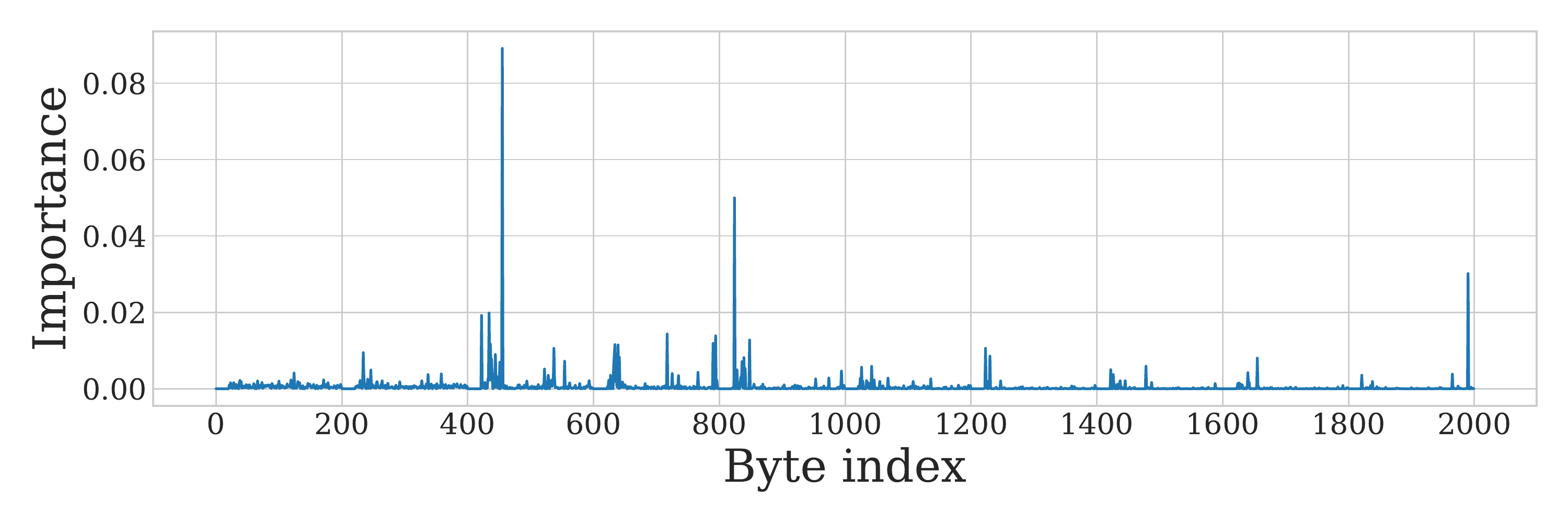}
			\label{fig:keeperImp3.1}
		}
		&
		\subfloat[The TLS Payloads with IV vectors masked]{%
			\includegraphics[width=0.45\textwidth]{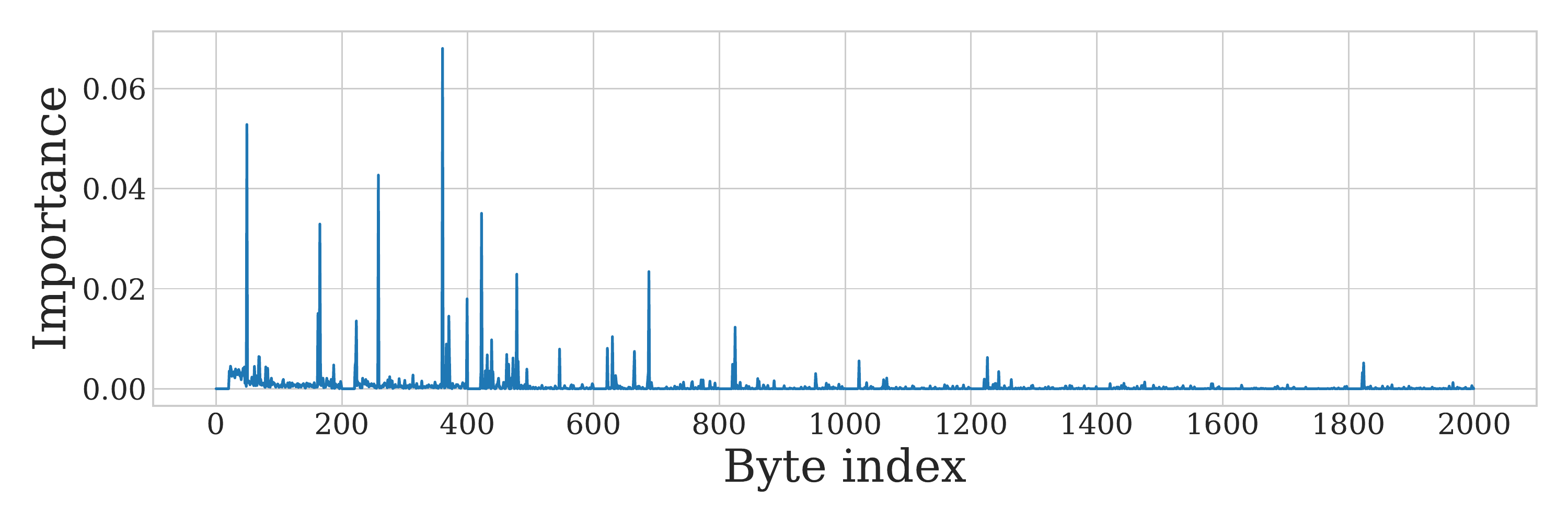}
			\label{fig:callerImp3.1}
		}
		\\
		\subfloat[The TLS Concatenated Payloads]{%
			\includegraphics[width=0.45\textwidth]{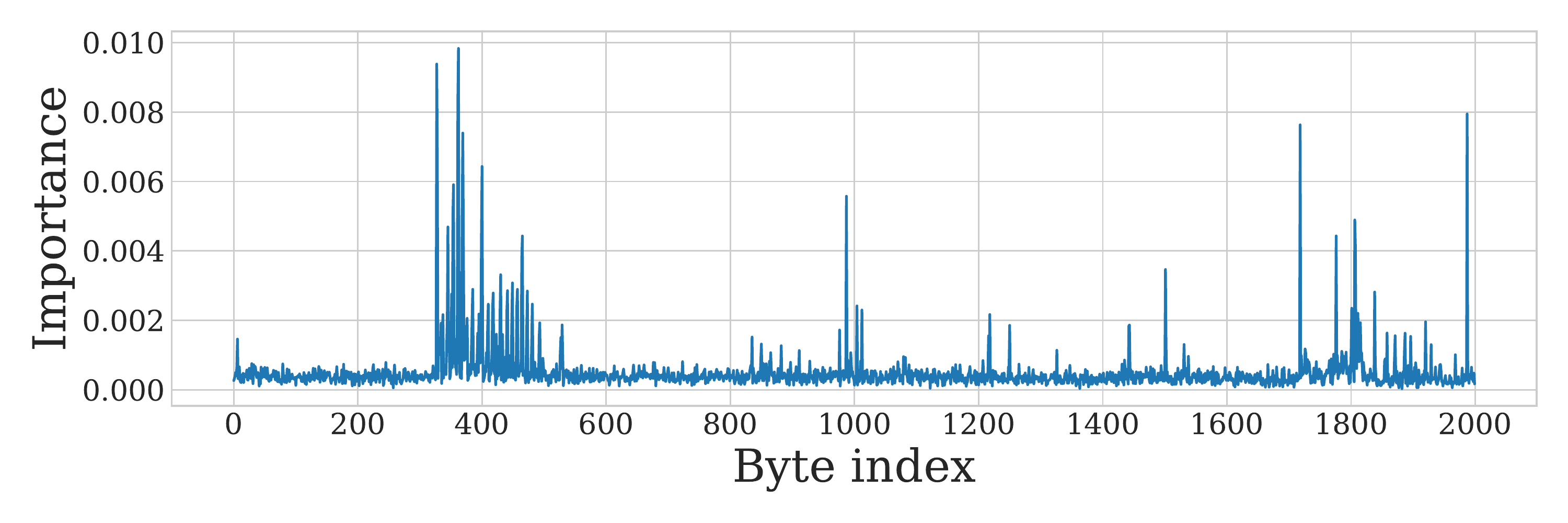}
			\label{fig:keeperImp4}
		}
		&
		\subfloat[The TLS Concatenated Payloads]{%
			\includegraphics[width=0.45\textwidth]{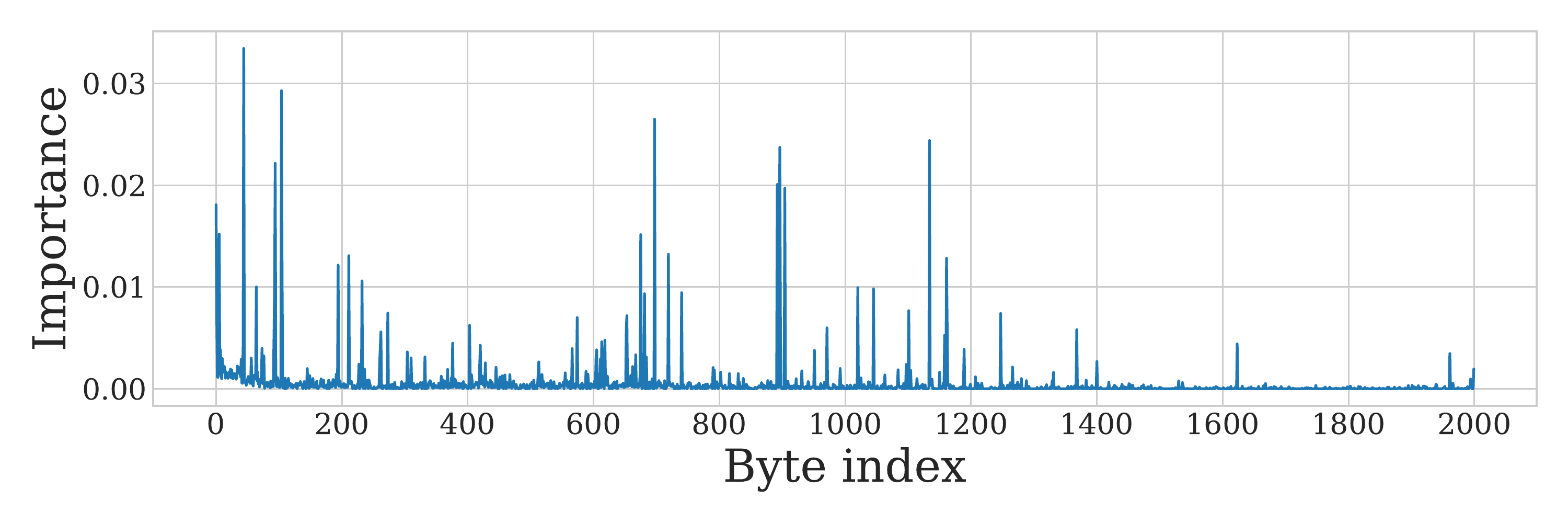}
			\label{fig:callerImp4}
		}
	\end{tabular}

\end{figure*}

The impact of input bytes on the output of the decision tree for both keeper and caller domain datasets are shown in Figs.~\ref{fig:imps}a and \ref{fig:imps}b. Considering the impact vectors, we can observe that the first and second packets have the most impact on the output. We know that the first TCP segments of a session conveying the TLS data contain TLS handshake records. The TLS handshakes mainly contain unencrypted data. Hence, we may conclude that the unencrypted TLS handshakes cause such high accuracy and information leakage. In the next iteration, we will set aside the handshake records and only focus on the TLS \emph{application data} records\footnote{Note that a TLS application data record is determined by the number 0x17 appearing in the first byte of the corresponding TCP payload.}.

\textbf{Second iteration: The TLS packets.} 
In the second iteration, as mentioned above, we only consider those TCP segments that contain TLS application data records. Similar to the previous iteration, we consider the first 200 bytes of the first 10 TLS application data packets in a given session as the input data. Then, we apply another iteration of Framework~\ref{fw:ProposedFramework} on this data. The result of the framework is reported in the Table~\ref{tab:accs}. We observe that the decision tree has an accuracy close enough to the accuracy of deep neural networks. So, we can rely on the interpretation result of the decision tree.

The impact of each byte on the output can be found in Figs. \ref{fig:imps}c and \ref{fig:imps}d. Considering these impact vectors, we can deduce that the 4th and 5th bytes of packets have the most impact on the output. Referring to the TLS protocol design, these bytes indicate the length of the TLS application data. So, we conclude that the TLS records' length is the most leaker part of the data in this iteration. In the next iteration, we use the TLS packets and mask their headers (\ie, the first five bytes of the packet) by setting their values to zero in order to remove the effect of these influential bytes. Notice that the first byte of the TLS application data contains the value 0x17 and the second and third bytes determine the protocol version. Since the first three bytes are not important, we make all the TLS header zero.

\textbf{Third iteration: The TLS packets with masked header.}
In the third iteration, we use the data used in the previous iteration (\ie, the TLS application data) with masked headers, namely, the first five bytes of such TLS packets are set to zero. Then we apply our ML methods on this data. The model accuracies of this scenario are also presented in Table~\ref{tab:accs}. Similar to the previous iterations, the decision tree has an accuracy close enough to the accuracies of deep neural networks. 

The impact of each byte on the output is depicted in Figs. \ref{fig:imps}e and \ref{fig:imps}f. Considering these impact vectors, one can observe that the most influential dimensions correspond to the initialization vector (IV) in the TLS protocol, which spans 6th to 21st bytes in the TLS application data payload.


\textbf{Fourth iteration: The TLS payloads with IV removed.}
In this iteration, we examine the effect of removing the encryption IV bytes in the TLS payload. Hence, to prepare the data for this iteration, we mask the first 21st bytes of each packet of the TLS application data. The model accuracies of this scenario are also presented in Table~\ref{tab:accs}. The results show that the performance of the decision tree is still close to the performance of deep neural networks. So the interpretation results of the decision tree are valid. Finally, the impact vectors of both datasets are visualized in Figs.~\ref{fig:imps}g and \ref{fig:imps}h.

\textbf{Fifth iteration: The Concatenated TLS packets.} 
The input data to this iteration is the data from the previous iteration, but with concatenated payloads. In other words, instead of picking first 200 bytes from encrypted sessions, we pick the whole payloads of the TLS data. Then, we link these payloads until we have a sequence of 2000 bytes. We apply our ML methods on such data. The hyperparameter values and model accuracies of this scenario are presented in the Appendix and Table~\ref{tab:accs}, respectively. The impact of each byte of this data is depicted in Figs.~\ref{fig:imps}i and \ref{fig:imps}j. Considering these impact vectors, we can observe that the impacts are not concentrated on some input dimensions anymore, and almost all input bytes have an impact less or more.

On the other hand, the accuracy of the decision tree is significantly less than neural networks, and the dependencies are so complex that the decision tree is not able to extract information as well as neural networks. So, based on our iterative algorithm, we can deduce that the process can be terminated since no specific part of data is significantly more leaker than the rest of the data. Also, all the data in this iteration are encrypted bytes, and it is almost impossible to continue our framework.

It is important to note that even at the end of our iterations, the information extractors' accuracy is much more than a random classifier. So, there still exists some information extractable in the TLS encrypted data, while it is not due to the leakage of information from specific bytes.

\begin{table*}\centering
	\caption{The accuracy of each ML model on the keeper and caller domain datasets (the numbers are in percent). Iteration 1: The TCP payloads, Iteration 2: The TLS packets, Iteration 3: The TLS packets with masked header, The Iteration 4: TLS payloads with IV removed, Iteration 5: Concatenated TLS packets.}
    \ra{1.3}
	\begin{tabular}{@{}lp{1cm}p{1cm}p{1cm}p{1cm}p{1.3cm}cp{1cm}p{1cm}p{1cm}p{1cm}p{1.3cm}@{}}\toprule
		\multirow{3}{*}{\textbf{Dataset}} & \multicolumn{5}{c}{\textbf{Keeper Domain Dataset}} & \phantom{abc}& \multicolumn{5}{c}{\textbf{Caller Domain Dataset}} \\ \cmidrule{2-6} \cmidrule{8-12}
		& \textbf{Itr. 1} & \textbf{Itr. 2} & \textbf{Itr. 3} & \textbf{Itr. 4} & \textbf{Itr. 5} && \textbf{Itr. 1} & \textbf{Itr. 2} & \textbf{Itr. 3} & \textbf{Itr. 4} & \textbf{Itr. 5}\\ \midrule
		\textbf{1-D Conv} & 98.19 & 49.30 & 27.44 & 25.48 & 21.71 && 60.26 & 29.55 & 26.58 & 19.85 & 20.16 \\
		\textbf{RNN} & 98.18 & 72.81 & 27.65 &  20.3 & 37.10 && 59.30 & 43.51 & 30.17 & 23.98 & 35.37 \\
		\textbf{DT}  & 98.58 & 72.01 & 25.21 & 19.1 & 15.57 && 63.25 & 46.62 & 25.95 & 20.00 & 24.80 \\
		\bottomrule
	\end{tabular}
    \label{tab:accs}
\end{table*}

%% file: 08-conclusion.tex
Interpretability of machine learning model is one of the interesting topics in the field of applied ML. In this paper, we have proposed a general framework that leverages the interpretation techniques from ML to analyze network protocols.
Then, as a demonstrating example, we have investigated the leakage of information from the TLS protocol, using our proposed framework. This approach helps us to understand the information leakage sources of the TLS traffic.

To this end, we first have implemented a website fingerprinting attack to the TLS protocol. We have gathered suitable datasets with the help of Selenium and Scrapy tools. Then, we have tried to extract information from traffic generated while a simulated user is surfing a website, using ML techniques. We have employed decision trees, recurrent neural networks, and convolutional neural networks to perform the fingerprinting attack. Moreover, for the interpreting technique, we use the importance vector provided by the decision tree algorithm. Our results show that the TLS handshake (which is mainly unencrypted), the TLS record length appearing in the TLS application data header, and the initialization vector (IV) field are among the most critical leaker parts in this protocol, respectively (also, see Fig.~\ref{fig:imps}).


Our results show that by eliminating the leaker dimensions (\ie, input bytes), the accuracy of the proposed fingerprinting models, used as the information extractor, decrease from 98.58\% to 15.57\% for the keeper domain setting and 63.25\% to 24.80\% for the caller domain setting, respectively. This decrement is significant, but there is still some information extractable from the encrypted data itself. Moreover, this information is not omittable since it is distributed almost evenly over all bytes of encrypted data (see Fig.~\ref{fig:imps}). So, it seems that the TLS protocol still needs some changes to get safer.

It should be emphasized that the proposed framework can be used for examining the information leakage from various protocols. Our framework suggests a reliable and systematic method for finding leaker parts of data, that can help protocol designers to make safer protocols.

As a future direction, instead of using the importance vector of decision trees for the interpretation method in our framework, one can employ interpretability techniques for deep learning, studied recently. However, it should be noted that although many techniques have been proposed to interpret DL models, this field is still under massive investigation. Moreover, many of the interpretation methods developed so far are mainly proposed for image applications.

Another interesting direction for future work is to study other types of information present in the Internet traffic data rather than the domain name. Of these information types include the content of the session, e.g., whether it is a video, a text message, an audio, or an image. It can be investigated  how much information is leaked from different content types and which parts of their sessions are the most responsible for this leakage. 
Distinguishing registered users of popular websites and online applications is yet another problem that is worth investigating. If there is some information leakage about the identity of registered users of a particular web service, it raises privacy concerns. 
Exploring these ideas is of great help to obtain more insight on the safety of existing Internet protocols.


%% file: 09-acknowledge.tex
The authors would like to thank Saeed Aqamiri for his valuable discussion and feedback on this work.

%% file: 10-appendix.tex
\subsection{Hyperparameters of the Proposed RNN}
The value of hyperparameters for the proposed RNN networks are listed in Table~\ref{tab:prop-lstm}.

\subsection{Hyperparameters of the Proposed CNN}
The value of hyperparameters for the proposed CNN network are listed in Table~\ref{tab:prop-cnn}.

\begin{table*}[t]\centering
	\ra{1.3}
	\begin{tabular}{@{}lp{1cm}p{1cm}p{1cm}p{1cm}p{1cm}cp{1cm}p{1cm}p{1cm}p{1cm}p{1cm}@{}}\toprule
		\multirow{3}{*}{\textbf{Dataset}} & \multicolumn{5}{c}{\textbf{Keeper Domain Dataset}} & \phantom{abc}& \multicolumn{5}{c}{\textbf{Caller Domain Dataset}} \\ \cmidrule{2-6} \cmidrule{8-12}
		& \textbf{Itr. 1} & \textbf{Itr. 2} & \textbf{Itr. 3} & \textbf{Itr. 4}& \textbf{Itr. 5} && \textbf{Itr. 1} & \textbf{Itr. 2} & \textbf{Itr. 3} & \textbf{Itr. 4} & \textbf{Itr. 5}\\ \midrule
		\textbf{Optimizer} &  adam & adam & adam & adam & adam & & adam & adam & adam & adam & adam \\
		\textbf{Learning Rate} & 0.001 & 0.001 & 0.001 & 0.001 & 0.001 &  & 0.001 & 0.001 & 0.001 & 0.001 & 0.001 \\
		\textbf{\# LSTM Layers} &  1 & 3 & 2 & 1 & 2 &   & 2 & 2 & 3 & 1 & 2 \\
		\textbf{\# LSTM Neurons} & [32] & [64, 16, 16] & [16, 16] & [32] & [64, 32] &   & [32, 16] & [16, 32] & [32, 32, 64] & [32] & [32, 32] \\
		\textbf{\# FC Layers} & 3 & 1 & 2 & 3 & 1 &   & 1 & 3 & 1 & 1 & 1 \\
		\textbf{\# FC Neurons} & [256, 256, 128] & [256] & [512, 256] & [256, 128, 256] & [256] &   & [128] & [256, 512, 128] & [512] & [512] & [256] \\
		\textbf{Activation Function}  & relu & tanh & tanh & relu & sotfplus &   & softplus & relu & relu & relu & tanh\\
		\textbf{Dropout Probability}  & 0.3 & 0.03 & 0.16 & 0.29 & 0.11 &   & 0.17 & 0.02 & 0.24 & 0.36 & 0.05 \\
		\bottomrule
	\end{tabular}
	\caption{The final values of hyperparameter for the RNN networks, used in Section~\ref{sec:results}.}
	\label{tab:prop-lstm}
\end{table*}

\begin{table*}[t]\centering
	\ra{1.3}
	\begin{tabular}{@{}lp{1cm}p{1cm}p{1cm}p{1cm}p{1cm}cp{1cm}p{1cm}p{1cm}p{1cm}p{1cm}@{}}\toprule
		\multirow{3}{*}{\textbf{Dataset}} & \multicolumn{5}{c}{\textbf{Keeper Domain Dataset}} & \phantom{abc}& \multicolumn{5}{c}{\textbf{Caller Domain Dataset}} \\ \cmidrule{2-6} \cmidrule{8-12}
		& \textbf{Itr. 1} & \textbf{Itr. 2} & \textbf{Itr. 3} & \textbf{Itr. 4} & \textbf{Itr. 5} && \textbf{Itr. 1} & \textbf{Itr. 2} & \textbf{Itr. 3} & \textbf{Itr. 4} & \textbf{Itr. 5}\\ \midrule
		\textbf{Optimizer} &  adam & adam & adam & adam & adam & & adam & adam & adam & x & adam\\
		\textbf{Learning Rate} & 0.0001 & 0.0001 & 0.0001 & 0.0001 & 0.0001 &  & 0.001 & 0.0001 & 0.0001 & 0.0001 & 0.001 \\
		\textbf{Stride} & 1 & 4 & 1 & 8 & 4 &  & 4 & 1 & 1 & 1 & 8 \\
		\textbf{\# Conv Layers} &  1 & 2 & 1 & 1 & 1 &   & 1 & 2 & 3 & x & 3 \\
		\textbf{\# Filters} & [128] & [256, 256] &[64] & [128] & [256] &   & [64] & [64, 256] & [128, 64, 256] & [64, 256] & [64, 64, 128]\\
		\textbf{Kernel} & [32] & [8, 128] & [32] & [128] & [32] &  & [8] & [8, 32] & [128, 32, 32] & [[32, 128]] & [32, 32, 8] \\
		\textbf{\# FC Layers} & 3 & 1 & 3 & 1 & 2 &  & 2 & 3 & 2 & 2 & 3 \\
		\textbf{\# FC Neurons} & [256, 512, 512] & [256] & [256, 128, 512] & [128] & [512, 256] &  & [512, 128] & [512, 512, 128] & [128, 128] & [128, 128] & [128, 256, 256]\\
		\textbf{Activation Function}  & softplus & softplus & softplus & tanh & relu &   & relu & tanh & softplus & tanh & tanh \\
		\textbf{Dropout Probability}  & 0.44 & 0.47 & 0.16 & 0.08 & 0.37 &  & 0.17 & 0.34 & 0.09 & 0.49 & 0.15 \\
		\bottomrule
	\end{tabular}
	\caption{The final values of hyperparameter for the CNN networks, used in Section~\ref{sec:results}.}
	\label{tab:prop-cnn}
\end{table*}

%
%

%% file: webfp-v2.bbl
\begin{thebibliography}{60}
\providecommand{\natexlab}[1]{#1}
\providecommand{\url}[1]{{#1}}
\providecommand{\urlprefix}{URL }
\expandafter\ifx\csname urlstyle\endcsname\relax
  \providecommand{\doi}[1]{DOI~\discretionary{}{}{}#1}\else
  \providecommand{\doi}{DOI~\discretionary{}{}{}\begingroup
  \urlstyle{rm}\Url}\fi
\providecommand{\eprint}[2][]{\url{#2}}

\bibitem[{Sel(2020)}]{Selenium}
 (2020) A browser automation framework and ecosystem.
  https://github.com/SeleniumHQ/selenium

\bibitem[{nam(2020)}]{namespace}
 (2020) namespace(7) - linux manual page.
  https://man7.org/linux/man-pages/man7/namespaces.7.html

\bibitem[{Scr(2020)}]{Scrapy}
 (2020) Scrapy, a fast high-level web crawling \& scraping framework for
  python. https://github.com/scrapy/scrapy

\bibitem[{Aceto et~al(2019)Aceto, Ciuonzo, Montieri, and
  Pescap{\'e}}]{aceto2019mobile}
Aceto G, Ciuonzo D, Montieri A, Pescap{\'e} A (2019) Mobile encrypted traffic
  classification using deep learning: Experimental evaluation, lessons learned,
  and challenges. IEEE Transactions on Network and Service Management

\bibitem[{Anderson and McGrew(2019)}]{10.1145/3355369.3355601}
Anderson B, McGrew D (2019) Tls beyond the browser: Combining end host and
  network data to understand application behavior. In: Proceedings of the
  Internet Measurement Conference, Association for Computing Machinery, New
  York, NY, USA, IMC '19, pp 379--392, \doi{10.1145/3355369.3355601},
  \urlprefix\url{https://doi.org/10.1145/3355369.3355601}

\bibitem[{Aviram et~al(2016)Aviram, Schinzel, Somorovsky, Heninger, Dankel,
  Steube, Valenta, Adrian, Halderman, Dukhovni, K{\"a}sper, Cohney, Engels,
  Paar, and Shavitt}]{Aviram2016DROWNBT}
Aviram N, Schinzel S, Somorovsky J, Heninger N, Dankel M, Steube J, Valenta L,
  Adrian D, Halderman JA, Dukhovni V, K{\"a}sper E, Cohney SN, Engels S, Paar
  C, Shavitt Y (2016) Drown: Breaking tls using sslv2. In: USENIX Security
  Symposium

\bibitem[{Bakhshi and Ghita(2016)}]{bakhshi2016internet}
Bakhshi T, Ghita B (2016) On internet traffic classification: A two-phased
  machine learning approach. Journal of Computer Networks and Communications
  2016

\bibitem[{Barnes et~al(2015)Barnes, Thomson, Pironti, and
  Langley}]{barnes2015deprecating}
Barnes R, Thomson M, Pironti A, Langley A (2015) Deprecating secure sockets
  layer version 3.0. In: IETF RFC 7568

\bibitem[{Bergstra et~al(2011)Bergstra, Bardenet, Bengio, and
  K{\'e}gl}]{bergstra2011algorithms}
Bergstra JS, Bardenet R, Bengio Y, K{\'e}gl B (2011) Algorithms for
  hyper-parameter optimization. In: Advances in neural information processing
  systems, pp 2546--2554

\bibitem[{Bermolen et~al(2011)Bermolen, Mellia, Meo, Rossi, and
  Valenti}]{bermolen2011abacus}
Bermolen P, Mellia M, Meo M, Rossi D, Valenti S (2011) Abacus: Accurate
  behavioral classification of p2p-tv traffic. Computer Networks
  55(6):1394--1411

\bibitem[{Bernaille et~al(2006)Bernaille, Teixeira, Akodkenou, Soule, and
  Salamatian}]{Bernaille2006}
Bernaille L, Teixeira R, Akodkenou I, Soule A, Salamatian K (2006) Traffic
  classification on the fly. SIGCOMM Comput Commun Rev 36(2):23--26,
  \doi{10.1145/1129582.1129589},
  \urlprefix\url{http://doi.acm.org/10.1145/1129582.1129589}

\bibitem[{B\"{o}ck et~al(2018)B\"{o}ck, Somorovsky, and
  Young}]{10.5555/3277203.3277265}
B\"{o}ck H, Somorovsky J, Young C (2018) Return of bleichenbacher's oracle
  threat (robot). In: Proceedings of the 27th USENIX Conference on Security
  Symposium, USENIX Association, USA, SEC'18, pp 817--832

\bibitem[{Boutaba et~al(2018)Boutaba, Salahuddin, Limam, Ayoubi, Shahriar,
  Estrada-Solano, and Caicedo}]{boutaba2018comprehensive}
Boutaba R, Salahuddin MA, Limam N, Ayoubi S, Shahriar N, Estrada-Solano F,
  Caicedo OM (2018) A comprehensive survey on machine learning for networking:
  evolution, applications and research opportunities. Journal of Internet
  Services and Applications 9(1):16

\bibitem[{Bujlow et~al(2015)Bujlow, Carela-Espa{\~n}ol, and
  Barlet-Ros}]{bujlow2015independent}
Bujlow T, Carela-Espa{\~n}ol V, Barlet-Ros P (2015) Independent comparison of
  popular dpi tools for traffic classification. Computer Networks 76:75--89

\bibitem[{Cai et~al(2012)Cai, Zhang, Joshi, and Johnson}]{cai2012touching}
Cai X, Zhang XC, Joshi B, Johnson R (2012) Touching from a distance: Website
  fingerprinting attacks and defenses. In: Proceedings of the 2012 ACM
  Conference on Computer and Communications Security, ACM, New York, NY, USA,
  CCS '12, pp 605--616, \doi{10.1145/2382196.2382260},
  \urlprefix\url{http://doi.acm.org/10.1145/2382196.2382260}

\bibitem[{Deri et~al(2014)Deri, Martinelli, Bujlow, and
  Cardigliano}]{deri2014ndpi}
Deri L, Martinelli M, Bujlow T, Cardigliano A (2014) ndpi: Open-source
  high-speed deep packet inspection. In: 2014 International Wireless
  Communications and Mobile Computing Conference (IWCMC), IEEE, pp 617--622

\bibitem[{Dierks and Rescorla(2008)}]{dierks2008transport}
Dierks T, Rescorla E (2008) The transport layer security (tls) protocol version
  1.2

\bibitem[{Draper-Gil et~al(2016)Draper-Gil, Lashkari, Mamun, and
  Ghorbani}]{draper2016characterization}
Draper-Gil G, Lashkari AH, Mamun MSI, Ghorbani AA (2016) Characterization of
  encrypted and vpn traffic using time-related. In: Proceedings of the 2nd
  international conference on information systems security and privacy
  (ICISSP), pp 407--414

\bibitem[{Erman et~al(2007)Erman, Mahanti, Arlitt, and
  Williamson}]{erman2007identifying}
Erman J, Mahanti A, Arlitt M, Williamson C (2007) Identifying and
  discriminating between web and peer-to-peer traffic in the network core. In:
  Proceedings of the 16th international conference on World Wide Web, ACM, pp
  883--892

\bibitem[{Frolov and Wustrow(2019)}]{Frolov2019TheUO}
Frolov S, Wustrow E (2019) The use of tls in censorship circumvention. In: NDSS

\bibitem[{Goodfellow et~al(2016)Goodfellow, Bengio, and
  Courville}]{Goodfellow-et-al-2016}
Goodfellow I, Bengio Y, Courville A (2016) Deep Learning. MIT Press,
  \url{http://www.deeplearningbook.org}

\bibitem[{Graves(2012{\natexlab{a}})}]{graves2012long}
Graves A (2012{\natexlab{a}}) Long short-term memory. In: Supervised sequence
  labelling with recurrent neural networks, Springer, pp 37--45

\bibitem[{Graves(2012{\natexlab{b}})}]{graves2012supervised}
Graves A (2012{\natexlab{b}}) Supervised sequence labelling. In: Supervised
  sequence labelling with recurrent neural networks, Springer, pp 5--13

\bibitem[{Hastie et~al(2009)Hastie, Tibshirani, and
  Friedman}]{hastie2009elements}
Hastie T, Tibshirani R, Friedman J (2009) The elements of statistical learning:
  data mining, inference, and prediction. Springer Science \& Business Media

\bibitem[{Holz et~al(2016)Holz, Amann, Mehani, Wachs, and
  Ali~Kaafar}]{Holz_2016}
Holz R, Amann J, Mehani O, Wachs M, Ali~Kaafar M (2016) Tls in the wild: An
  internet-wide analysis of tls-based protocols for electronic communication.
  Proceedings 2016 Network and Distributed System Security Symposium
  \doi{10.14722/ndss.2016.23055},
  \urlprefix\url{http://dx.doi.org/10.14722/ndss.2016.23055}

\bibitem[{Hu et~al(2011)Hu, Che, Zhang, Zhang, Guo, and Yu}]{hu2011rank}
Hu Q, Che X, Zhang L, Zhang D, Guo M, Yu D (2011) Rank entropy-based decision
  trees for monotonic classification. IEEE Transactions on Knowledge and Data
  Engineering 24(11):2052--2064

\bibitem[{Hus\'{a}k et~al(2015)Hus\'{a}k, Cerm\'{a}k, Jirs\'{\i}k, and
  Celeda}]{10.1109/ARES.2015.35}
Hus\'{a}k M, Cerm\'{a}k M, Jirs\'{\i}k T, Celeda P (2015) Network-based https
  client identification using ssl/tls fingerprinting. In: Proceedings of the
  2015 10th International Conference on Availability, Reliability and Security,
  IEEE Computer Society, USA, ARES '15, pp 389--396,
  \doi{10.1109/ARES.2015.35},
  \urlprefix\url{https://doi.org/10.1109/ARES.2015.35}

\bibitem[{Hus\'{a}k et~al(2016)Hus\'{a}k, \v{C}erm\'{a}k, Jirs\'{\i}k, and
  \v{C}eleda}]{10.1186/s13635-016-0030-7}
Hus\'{a}k M, \v{C}erm\'{a}k M, Jirs\'{\i}k T, \v{C}eleda P (2016) Https traffic
  analysis and client identification using passive ssl/tls fingerprinting.
  EURASIP J Inf Secur 2016(1), \doi{10.1186/s13635-016-0030-7},
  \urlprefix\url{https://doi.org/10.1186/s13635-016-0030-7}

\bibitem[{Ioffe and Szegedy(2015)}]{ioffe2015batch}
Ioffe S, Szegedy C (2015) Batch normalization: Accelerating deep network
  training by reducing internal covariate shift. In: International conference
  on machine learning, pp 448--456

\bibitem[{Karagiannis et~al(2005)Karagiannis, Papagiannaki, and
  Faloutsos}]{karagiannis2005blinc}
Karagiannis T, Papagiannaki K, Faloutsos M (2005) Blinc: multilevel traffic
  classification in the dark. In: ACM SIGCOMM Computer Communication Review,
  ACM, vol~35, pp 229--240

\bibitem[{Kim et~al(2008)Kim, Claffy, Fomenkov, Barman, Faloutsos, and
  Lee}]{kim2008internet}
Kim H, Claffy KC, Fomenkov M, Barman D, Faloutsos M, Lee K (2008) Internet
  traffic classification demystified: myths, caveats, and the best practices.
  In: Proceedings of the 2008 ACM CoNEXT conference, ACM, p~11

\bibitem[{Kim(2014)}]{kim2014convolutional}
Kim Y (2014) Convolutional neural networks for sentence classification. arXiv
  preprint arXiv:14085882

\bibitem[{Kotzias et~al(2018)Kotzias, Razaghpanah, Amann, Paterson,
  Vallina-Rodriguez, and Caballero}]{10.1145/3278532.3278568}
Kotzias P, Razaghpanah A, Amann J, Paterson KG, Vallina-Rodriguez N, Caballero
  J (2018) Coming of age: A longitudinal study of tls deployment. In:
  Proceedings of the Internet Measurement Conference 2018, Association for
  Computing Machinery, New York, NY, USA, IMC '18, pp 415--428,
  \doi{10.1145/3278532.3278568},
  \urlprefix\url{https://doi.org/10.1145/3278532.3278568}

\bibitem[{Lin et~al(2009)Lin, Lu, Lai, Peng, and Lin}]{LIN2009packetsize}
Lin YD, Lu CN, Lai YC, Peng WH, Lin PC (2009) Application classification using
  packet size distribution and port association. Journal of Network and
  Computer Applications 32(5):1023 -- 1030,
  \doi{https://doi.org/10.1016/j.jnca.2009.03.001},
  \urlprefix\url{http://www.sciencedirect.com/science/article/pii/S1084804509000484},
  next Generation Content Networks

\bibitem[{Lopez-Martin et~al(2017)Lopez-Martin, Carro, Sanchez-Esguevillas, and
  Lloret}]{lopez2017network}
Lopez-Martin M, Carro B, Sanchez-Esguevillas A, Lloret J (2017) Network traffic
  classifier with convolutional and recurrent neural networks for internet of
  things. IEEE Access 5:18042--18050

\bibitem[{Lotfollahi et~al(2020)Lotfollahi, Siavoshani, Zade, and
  Saberian}]{lotfollahi2017deep}
Lotfollahi M, Siavoshani MJ, Zade RSH, Saberian M (2020) Deep packet: A novel
  approach for encrypted traffic classification using deep learning. Soft
  Computing

\bibitem[{M{\"o}ller et~al(2014)M{\"o}ller, Duong, and
  Kotowicz}]{moller2014poodle}
M{\"o}ller B, Duong T, Kotowicz K (2014) This poodle bites: exploiting the ssl
  3.0 fallback. Security Advisory 21:34--58

\bibitem[{Naylor et~al(2014)Naylor, Finamore, Leontiadis, Grunenberger, Mellia,
  Munafo, Papagiannaki, and Steenkiste}]{naylor2014cost}
Naylor D, Finamore A, Leontiadis I, Grunenberger Y, Mellia M, Munafo M,
  Papagiannaki K, Steenkiste P (2014) The cost of the `s' in https. In:
  Proceedings of the 10th ACM International Conference on emerging Networking
  Experiments and Technologies, pp 133--140

\bibitem[{Nikiforakis et~al(2013)Nikiforakis, Kapravelos, Joosen, Kruegel,
  Piessens, and Vigna}]{Nikiforakis2013cookie}
Nikiforakis N, Kapravelos A, Joosen W, Kruegel C, Piessens F, Vigna G (2013)
  Cookieless monster: Exploring the ecosystem of web-based device
  fingerprinting. pp 541--555, \doi{10.1109/SP.2013.43}

\bibitem[{Pacheco et~al(2018)Pacheco, Exposito, Gineste, Baudoin, and
  Aguilar}]{Pacheco2019TowardsTD}
Pacheco F, Exposito E, Gineste M, Baudoin C, Aguilar J (2018) Towards the
  deployment of machine learning solutions in network traffic classification: A
  systematic survey. IEEE Communications Surveys \& Tutorials 21(2):1988--2014

\bibitem[{Panchenko et~al(2016)Panchenko, Lanze, Pennekamp, Engel, Zinnen,
  Henze, and Wehrle}]{panchenko2016website}
Panchenko A, Lanze F, Pennekamp J, Engel T, Zinnen A, Henze M, Wehrle K (2016)
  Website fingerprinting at internet scale. In: NDSS

\bibitem[{Quinlan(1986)}]{quinlan1986induction}
Quinlan JR (1986) Induction of decision trees. Machine learning 1(1):81--106

\bibitem[{Razaghpanah et~al(2017)Razaghpanah, Akhavan~Niaki, Vallina-Rodriguez,
  Sundaresan, Amann, and Gill}]{tls-android}
Razaghpanah A, Akhavan~Niaki A, Vallina-Rodriguez N, Sundaresan S, Amann J,
  Gill P (2017) Studying tls usage in android apps. pp 350--362,
  \doi{10.1145/3143361.3143400}

\bibitem[{Rezaei and Liu(2019)}]{rezaei2019deep}
Rezaei S, Liu X (2019) Deep learning for encrypted traffic classification: An
  overview. IEEE communications magazine 57(5):76--81

\bibitem[{Rimmer et~al(2017)Rimmer, Preuveneers, Ju{\'{a}}rez, van Goethem, and
  Joosen}]{rimmer2017AutFeatureExtraction}
Rimmer V, Preuveneers D, Ju{\'{a}}rez M, van Goethem T, Joosen W (2017)
  Automated feature extraction for website fingerprinting through deep
  learning. CoRR abs/1708.06376,
  \urlprefix\url{http://arxiv.org/abs/1708.06376}, \eprint{1708.06376}

\bibitem[{Rimmer et~al(2018)Rimmer, Preuveneers, Juarez, Goethem, and
  Joosen}]{Rimmer2018automated}
Rimmer V, Preuveneers D, Juarez M, Goethem TV, Joosen W (2018) Automated
  website fingerprinting through deep learning. Proceedings 2018 Network and
  Distributed System Security Symposium \doi{10.14722/ndss.2018.23105},
  \urlprefix\url{http://dx.doi.org/10.14722/ndss.2018.23105}

\bibitem[{Roughan et~al(2004)Roughan, Sen, Spatscheck, and
  Duffield}]{roughan2004class}
Roughan M, Sen S, Spatscheck O, Duffield N (2004) Class-of-service mapping for
  qos: a statistical signature-based approach to ip traffic classification. In:
  Proceedings of the 4th ACM SIGCOMM conference on Internet measurement, ACM,
  pp 135--148

\bibitem[{Schmidhuber(2015)}]{schmidhuber2015deep}
Schmidhuber J (2015) Deep learning in neural networks: An overview. Neural
  networks 61:85--117

\bibitem[{Shen et~al(2019)Shen, Liu, Chen, Zhu, and Zhang}]{shen2019webpage}
Shen M, Liu Y, Chen S, Zhu L, Zhang Y (2019) Webpage fingerprinting using only
  packet length information. In: ICC 2019-2019 IEEE International Conference on
  Communications (ICC), IEEE, pp 1--6

\bibitem[{Sherry et~al(2015)Sherry, Lan, Popa, and
  Ratnasamy}]{sherry2015blindbox}
Sherry J, Lan C, Popa RA, Ratnasamy S (2015) Blindbox: Deep packet inspection
  over encrypted traffic. In: ACM SIGCOMM Computer Communication Review, ACM,
  vol~45, pp 213--226

\bibitem[{Sirinam et~al(2018)Sirinam, Imani, Ju{\'{a}}rez, and
  Wright}]{sirinam2018deep}
Sirinam P, Imani M, Ju{\'{a}}rez M, Wright M (2018) Deep fingerprinting:
  Undermining website fingerprinting defenses with deep learning. CoRR
  abs/1801.02265, \urlprefix\url{http://arxiv.org/abs/1801.02265},
  \eprint{1801.02265}

\bibitem[{Soltani et~al(2020)Soltani, Siavoshani, and
  Jahangir}]{Soltani-DID-ArXiv20}
Soltani M, Siavoshani MJ, Jahangir AH (2020) A content-based deep intrusion
  detection system. CoRR abs/2001.05009,
  \urlprefix\url{https://arxiv.org/abs/2001.05009}

\bibitem[{Soltani et~al(2021)Soltani, Ousat, Siavoshani, and
  Jahangir}]{Soltani-adaptable-ArXiv21}
Soltani M, Ousat B, Siavoshani MJ, Jahangir AH (2021) An adaptable deep
  learning-based intrusion detection system to zero-day attacks. CoRR
  abs/2108.09199, \urlprefix\url{https://arxiv.org/abs/2108.09199}

\bibitem[{Stallings and Brown(2014)}]{Stallings2014security}
Stallings W, Brown L (2014) Computer Security: Principles and Practice, 3rd
  edn. Prentice Hall Press, Upper Saddle River, NJ, USA

\bibitem[{Verleysen and Fran{\c{c}}ois(2005)}]{verleysen2005curse}
Verleysen M, Fran{\c{c}}ois D (2005) The curse of dimensionality in data mining
  and time series prediction. In: International Work-Conference on Artificial
  Neural Networks, Springer, pp 758--770

\bibitem[{Wang et~al(2017{\natexlab{a}})Wang, Zhu, Wang, Zeng, and
  Yang}]{wang2017endtoend}
Wang W, Zhu M, Wang J, Zeng X, Yang Z (2017{\natexlab{a}}) End-to-end encrypted
  traffic classification with one-dimensional convolution neural networks. In:
  Intelligence and {Security} {Informatics} ({ISI}), 2017 {IEEE}
  {International} {Conference} on, IEEE, pp 43--48

\bibitem[{Wang et~al(2017{\natexlab{b}})Wang, Zhu, Zeng, Ye, and
  Sheng}]{wang2017malware}
Wang W, Zhu M, Zeng X, Ye X, Sheng Y (2017{\natexlab{b}}) Malware traffic
  classification using convolutional neural network for representation
  learning. In: 2017 International Conference on Information Networking
  (ICOIN), IEEE, pp 712--717

\bibitem[{Yen et~al(2012)Yen, Xie, Yu, Yu, and Abadi}]{yen2012host}
Yen TF, Xie Y, Yu F, Yu RP, Abadi M (2012) Host fingerprinting and tracking on
  the web: Privacy and security implications. In: NDSS, vol~62, p~66

\bibitem[{{Yufeng Kou} et~al(2004){Yufeng Kou}, {Chang-Tien Lu},
  {Sirwongwattana}, and {Yo-Ping Huang}}]{kou2004survey}
{Yufeng Kou}, {Chang-Tien Lu}, {Sirwongwattana} S, {Yo-Ping Huang} (2004)
  Survey of fraud detection techniques. In: IEEE International Conference on
  Networking, Sensing and Control, 2004, vol~2, pp 749--754 Vol.2,
  \doi{10.1109/ICNSC.2004.1297040}

\bibitem[{Zliobaite et~al(2016)Zliobaite, Pechenizkiy, and
  Gama}]{zliobaite2016}
Zliobaite I, Pechenizkiy M, Gama J (2016) An overview of concept drift
  applications. In: Big data analysis: new algorithms for a new society,
  Springer, pp 91--114

\end{thebibliography}
